\documentclass[twocolumn,twocolappendix]{aastex631}

\usepackage[utf8]{inputenc}
\usepackage[T1]{fontenc}
\usepackage{lmodern}
\usepackage[english]{babel}

\usepackage{amssymb,amsmath,amsfonts}
\usepackage{graphicx}
\usepackage{xparse}
\usepackage{xstring}
\usepackage{booktabs}
\usepackage{paralist}
\usepackage{enumitem}
\usepackage[all]{nowidow}

\usepackage{nameref}
\usepackage[capitalise]{cleveref} 
\usepackage{url} 

\usepackage{savesym}
\savesymbol{tablenum}
\usepackage{siunitx}
\restoresymbol{SIX}{tablenum}

\renewcommand{\vec}[1]{\mbox{\boldmath$#1$}}

\renewcommand{\arraystretch}{1.2}

\DeclareSIUnit \parsec {pc}
\DeclareSIUnit \eV {eV}
\DeclareSIUnit \eVcc {eV/$c^2$}
\DeclareSIUnit \Msun {M_\odot}

\setlist[description]{
  leftmargin=\parindent
}

\def \L {\mathbb{L}}
\def \N {\mathbb{N}}										
\def \Z {\mathbb{Z}}										
\def \Q {\mathbb{Q}}										
\def \R {\mathbb{R}}										
\def \C {\mathbb{C}}										

\def \E			#1{\cdot 10^{#1}}					
\def \e			{\mathrm{e}}							
\def \d			{\mathrm{d}}							
\def \i			{\mathfrak{i}}						
\def \const {\mathrm{const}}					

\newcommand{\degree}{\ensuremath{^\circ}}

\DeclareMathOperator{\diag}{diag}

\newcommand*\pFq[2]{{}_{#1}F_{#2}} 

\def \bracket		#1{\left(#1\right)}										
\def \abs				#1{\left|#1\right|}										

\newcommand \qbracket[2][]{\left[\vphantom{#1}#2\right]}								

\def \diff				#1#2{\frac{\d#1}{\d#2}}

\def \diffpart		#1#2{\frac{\partial #1}{\partial #2}}

\NewDocumentCommand{\DEFradius}{O{1}}{%
	\ifnum #1>0 %
		\frac{r^{#1}}{R^{#1}} %
	\else %
		\ifnum #1=-1 %
			\frac{R}{r} %
		\else %
			\frac{R^{\StrGobbleLeft{#1}{1}}}{r^{\StrGobbleLeft{#1}{1}}} %
		\fi %
	\fi %
}

\newcommand \SCLradiusB {R_{\text{\tiny B}}}

\NewDocumentCommand{\DEFradiusB}{O{1}}{%
	\ifnum #1>0 %
		\frac{r^{#1}}{\SCLradiusB^{#1}} %
	\else %
		\ifnum #1=-1 %
			\frac{\SCLradiusB}{r} %
		\else %
			\frac{\SCLradiusB^{\StrGobbleLeft{#1}{1}}}{r^{\StrGobbleLeft{#1}{1}}} %
		\fi %
	\fi %
}

\def \SCLfermiEnergy{\varepsilon_{\text{\tiny F}}}

\NewDocumentCommand{\DEFfermiEnergy}{O{1}}{%
	\ifnum #1=1 %
		\SCLfermiEnergy(r) %
	\else %
		\SCLfermiEnergy^{#1}(r) %
	\fi %
}

\def \SYMvbulge	{V_\mathrm{b}}
\def \SYMvdisk	{V_\mathrm{d}}
\def \SYMvgas	{V_\mathrm{g}}
\def \SYMvtot	{V_\mathrm{tot}}
\def \SYMvbar	{V_\mathrm{bar}}
\def \SYMvdark	{V_\mathrm{DM}} 

\def \SYMacc	{a}
\def \SYMatot	{a_\mathrm{tot}}

\def \SYMabar	{a_\mathrm{bar}}
\def \SYMaDM	{a_\mathrm{DM}}
\def \SYMafrak	{\mathfrak{a}_0}

\shorttitle{Galaxy rotation curves and universal scaling relations}
\shortauthors{Krut et al.}

\begin{document}

\title{Galaxy rotation curves and universal scaling relations: comparison between phenomenological and fermionic dark matter profiles}
\correspondingauthor{A. Krut, C. R. Argüelles}
\email{andreas.krut@icranet.org, carguelles@fcaglp.unlp.edu.ar}

\author{A. Krut}
\affiliation{ICRANet, Piazza della Repubblica 10, I-65122 Pescara, Italy}

\author{C. R. Argüelles}
\affiliation{ICRANet, Piazza della Repubblica 10, I-65122 Pescara, Italy}
\affiliation{Instituto de Astrof\'isica de La Plata, UNLP-CONICET, Paseo del Bosque s/n B1900FWA La Plata, Argentina}

\author{P.-H. Chavanis}
\affiliation{Laboratoire de Physique Th\'eorique, Universit\'e de Toulouse, CNRS, UPS, France}

\author{J. A. Rueda}
\affiliation{ICRANet, Piazza della Repubblica 10, I-65122 Pescara, Italy}
\affiliation{Dip. di Fisica, Sapienza Università di Roma, Piazzale Aldo Moro 5, I-00185 Rome, Italy}
\affiliation{ICRANet-Ferrara, Dip. di Fisica e Scienze della Terra, Universit\`a degli Studi di Ferrara, Via Saragat 1, 44122 Ferrara, Italy}
\affiliation{INAF, Istituto de Astrofisica e Planetologia Spaziali, Via Fosso del Cavaliere 100, 00133 Rome, Italy}

\author{R. Ruffini}
\affiliation{ICRANet, Piazza della Repubblica 10, I-65122 Pescara, Italy}
\affiliation{Dip. di Fisica, Sapienza Università di Roma, Piazzale Aldo Moro 5, I-00185 Rome, Italy}
\affiliation{ICRANet-Ferrara, Dip. di Fisica e Scienze della Terra, Universit\`a degli Studi di Ferrara, Via Saragat 1, 44122 Ferrara, Italy}
\affiliation{INAF, Istituto de Astrofisica e Planetologia Spaziali, Via Fosso del Cavaliere 100, 00133 Rome, Italy}
\begin{abstract}
Galaxies show different halo scaling relations such as the Radial Acceleration Relation, the Mass Discrepancy Acceleration Relation (MDAR) or the dark matter Surface Density Relation (SDR). At difference with traditional studies using phenomenological $\Lambda$CDM halos, we analyze the above relations assuming that dark matter (DM) halos are formed through a Maximum Entropy Principle (MEP) in which the fermionic (quantum) nature of the DM particles is dully accounted for. For the first time a competitive DM model based on first physical principles, such as (quantum) statistical-mechanics and thermodynamics, is tested against a large data-set of galactic observables.
In particular, we compare the fermionic DM model with empirical DM profiles: the NFW model, a generalized NFW model accounting for baryonic feedback, the Einasto model and the Burkert model. For this task, we use a large sample of $120$ galaxies taken from the Spitzer Photometry and Accurate Rotation Curves (SPARC) data-set, from which we infer the DM content to compare with the models. We find that the Radial Acceleration Relation and MDAR are well explained by all the models with comparable accuracy, while the fits to the individual rotation curves, in contrast, show that cored DM halos are statistically preferred with respect to the cuspy NFW profile.
However, very different physical principles justify the flat inner halo slope in the most favored DM profiles: while generalized NFW or Einasto models rely on complex baryonic feedback processes, the MEP scenario involves a quasi-thermodynamic equilibrium of the DM particles.
\end{abstract}


\section{Introduction}

How the total gravitating mass distributes with respect to the luminous mass on galaxy scales is an open question which has regained much attention in the last decade thanks to the vast data-sets covering broader radial extents across different Hubble types \citep{2008AJ....136.2648D,2011MNRAS.413..813C,2016AJ....152..157L}.
Several universal relations exist between different pairs of structural galaxy parameters, which refer either \begin{inparaenum}[(i)]
    \item to the outer regions of galaxies such as the baryonic Tully-Fisher relation (BTFR) \citep{2000ApJ...533L..99M}, the DM Surface Density Relation (SDR) \citep{2009MNRAS.397.1169D}, the Radial Acceleration Relation \citep{2016PhRvL.117t1101M}, and the Mass Discrepancy Acceleration Relation (MDAR) \citep{2004ApJ...609..652M}, which are indeed all closely related \citep{2004ApJ...609..652M,2016arXiv161208857S,2018FoPh...48.1517S}; 
    \item to their central regions such as the $M$-$\sigma$ relation between the bulge's dispersion velocity and the central object mass \citep{2000ApJ...539L...9F}; or
    \item to a combination of both regimes such as the Ferrarese relation \citep{2002ApJ...578...90F,2011Natur.469..377K,2015ApJ...800..124B} between the total halo mass and its supermassive central object mass.
\end{inparaenum}

Actual attempts for a unified understanding of many of the above scaling relations are typically given in terms of phenomenological halos obtained from N-body simulations within $\Lambda$CDM (see e.g. \citealp{2017MNRAS.471.1841N,2017PhRvL.118p1103L,2017MNRAS.464.2419S,2018FoPh...48.1517S}).

However, when DM halos are formed through a MEP for collisionless systems of self-gravitating fermions \citep{1998MNRAS.300..981C,2015PhRvD..92l3527C,2020EPJP..135..290C,2021MNRAS.502.4227A}, it leaves place to novel theoretical predictions in the phenomenology of real galaxies \citep{2018PDU....21...82A,2019PDU....24..278A,2020A&A...641A..34B,2021MNRAS.505L..64B,2021MNRAS.502.4227A,2022MNRAS.511L..35A} such as:

\begin{asparaenum}[(1)]
    \item DM fermions with a finite-temperature can be in a diluted (Boltzmannian-like) regime or become semi-degenerate. The corresponding DM halos can be, respectively, King-like or may develop a dense and degenerate compact core at the center of such a halo \citep{2015PhRvD..92l3527C,2021MNRAS.502.4227A}. A fully relativistic model in which this more general \textit{core}--\textit{halo} profiles arise, is usually referred in the literature as the Ruffini-Argüelles-Rueda (RAR) model \citep{2015MNRAS.451..622R,2018PDU....21...82A,2019PDU....24..278A,2020A&A...641A..34B,2021MNRAS.505L..64B,2021MNRAS.502.4227A,2022IJMPD..3130002A,2022MNRAS.511L..35A}. In either case, these fermionic DM profiles are cored (i.e. develop an extended plateau on halo scales similar to the Burkert DM profile as shown in \cref{fig:profile-illustration-mep}), thereby not suffering from the core-cusp problem associated with the standard $\Lambda$CDM cosmology \citep{2017ARA&A..55..343B}. This cored feature seems to be a general conclusion reached for any DM profile which has reached a (quasi) thermodynamic equilibrium in cosmology \citep{2021MNRAS.504.2832S}.

    \item The relevance of the fermionic solutions with a degenerate DM core surrounded by a diluted halo (named from now on as \textit{core}--\textit{halo} profiles) imply different consequences: (a) the core might become so densely packed that above a threshold --- the critical mass --- the quantum pressure can not support it any longer against its own weight, leading to the gravitational core-collapse into a supermassive black hole (SMBH;  \citealp{2020EPJB...93..208A,2021MNRAS.502.4227A}). For DM particle masses of $\mathcal{O}(10)$ keV, this result provides, for large enough galaxies, a novel SMBH formation mechanism in the early Universe as proposed in \citet{2021MNRAS.502.4227A}; and (b) for core masses below its critical value, it exist a set of free parameters in the fermionic model, such that the quantum cores correlate with their outer halos \citep{2019PDU....24..278A} explaining the relation between the total mass $M_{\rm tot}$ of large enough galaxies with the (presumably) embedded BH mass $M_{\rm BH}$, i.e. the Ferrarese relation (see point (iii) above).

    \item For small enough galaxies, as in the case of typical dwarfs, the degenerate core cannot collapse towards a BH (i.e. the total mass of the galaxy is below the critical mass of collapse) and thus it remains in a \textit{core}--\textit{halo} state where the central nucleus still mimics the effects of a singularity --- with core masses in the range of Intermediate mass BHs --- while the outer halo explains the rotation curves (RCs) \citep{2019PDU....24..278A,2021MNRAS.502.4227A,2022IJMPD..3130002A}.
    
    \item In the more extreme case where the fermions are fully-degenerate (i.e. when they are treated under the $T \to 0$ approximation), the corresponding halos are polytropic and may be only applicable to dwarfs \citep{2015JCAP...01..002D}.
\end{asparaenum}

Moreover, it was recently demonstrated that fermionic halos obtained via this MEP mechanism can arise in a cosmological framework, and remain thermodynamically and dynamically stable during the life of the Universe \citep{2021MNRAS.502.4227A}. Indeed, it was there shown the self-consistency of the approach in the sense that the nature and mass of the DM particles involved in the linear matter power spectrum --- calculated in \citet{2021MNRAS.502.4227A} within a CLASS code for $\mathcal{O}(10)$ keV fermions --- are the very same building blocks at the basis of the virialized DM configurations with its inherent effects in the DM profiles. Even if the MEP has been applied in \citet{2021MNRAS.502.4227A} within a warm DM (WDM) cosmology --- and is used here for finite temperature fermions (see \ref{sec:model:rar}) --- this is not univocal in general. Other MEP such as the one used in \citet{2021MNRAS.504.2832S} has been also applied for classical (self-gravitating) particles within cold DM (CDM) cosmologies.

Thus, the main purpose of this work is to analyze most of the galaxy relations, as discussed above in (i)-(iii), together with a large set of observed RCs provided by the SPARC data. In a forthcoming work we will reconsider this data-set together with observations coming from the central regions of galaxies (e.g. supermassive BHs), to include the $M_{\rm BH} - M_{\rm tot}$ relation in the study.

We assume that DM halos are formed through a MEP in which the fermionic (quantum) nature of the DM particles is dully accounted for. This takes special interest since it is the first time a predictive model of this kind, i.e. based on first physical principles such as (quantum) statistical-mechanics and thermodynamics, is tested against a large set of galaxy observables while leading to a good agreement with observations. To see how well this fermionic model can reproduce the given observables, it will be compared with most of the commonly used DM models used in the literature. Further, we check which set of observed data can (or cannot) discriminate the goodness of the competing models.

For this task, we will first focus on the Radial Acceleration Relation --- a non-linear correlation between the radial acceleration caused by the total matter and the one generated by its baryonic component only (see section \ref{sec:result:ac}) --- and on the directly related MDAR.

The main motivation to start studying these acceleration relations is due to the intense debate they have generated in the past few years about their underlying physical origin. For instance, \citet{2016PhRvL.117t1101M} argues that the Radial Acceleration Relation can be explained by a fundamental acceleration constant $\mathfrak{a}_0$ in combination with a modification of Newtonian gravity and without the necessity of any DM. One potential explanation for {such a fundamental constant} comes from Modified Newtonian Dynamics (MOND) \citep{2015CaJPh..93..169K,2016arXiv160906642M,2016PhRvL.117t1101M,2018A&A...615A...3L}, which has been used to interpret it as evidence against the $\Lambda$CDM paradigm and in favor to the MOND theory. However, more recent studies dedicated to analyze this universal relation within the (Bayesian) posterior distributions on the acceleration scales of individual galaxies (across a large sample), have provided evidence against the existence of such a fundamental constant and in favour of $\mathfrak{a}_0$ to be an emergent magnitude (see e.g. \citealp{2020MNRAS.494.2875M} and references therein). On the other hand, it has been extensively shown that the Radial Acceleration Relation is consistent with the $\Lambda$CDM paradigm, as found either in hydrodynamical N-body simulations  \citep{2016MNRAS.456L.127D,2017MNRAS.471.1841N,2017PhRvL.118p1103L,2019MNRAS.485.1886D}, or from other more phenomenological (independent) study based on Universal Rotation Curves \citep{2018FoPh...48.1517S}.

Thus, in the first part of this work we will use the MEP approach for fermions to evaluate how competitive it is to reproduce the above relations with respect to other phenomenological DM halos. The galactic observables are taken from a filtered SPARC sample of $2369$ data points (corresponding to a total of $120$ galaxies) and then apply a non-linear least square statistical analysis in order to check the goodness of fit of: 
\begin{inparaenum}[(a)]
    \item the above fermionic DM halos --- either in the \textit{core}--\textit{halo} regime \citep{2019PDU....24..278A,2021MNRAS.502.4227A} or in the purely King-like one \citep{2021MNRAS.502.4227A};
    \item a baryonic feedback motivated halo model within $\Lambda$CDM according to \citet{2014MNRAS.441.2986D}, named here as DC14;
    \item the classical NFW model based on early numerical simulations \citep{1997ApJ...490..493N};
    \item the Einasto profile \citep{1989A&A...223...89E,2006AJ....132.2685M}, and
    \item the Burkert model \citep{1995ApJ...447L..25B}.
\end{inparaenum}

Outlining the structure of this paper, in section \ref{sec:fitting} we give a brief overview about the SPARC data set, data selection and fitting procedure. In section \ref{sec:dark-matter-models}, we describe the competing DM halo models considered above. In section \ref{sec:results}, we present our results on the Radial Acceleration Relation by performing a goodness of the fit for each DM model based on a filtered SPARC sample. In section \ref{sec:fermionic-halos}, we focus on fermionic halos and compare their morphology with the other DM models. Finally, in section \ref{sec:conclusion}, we give a brief summary and draw the conclusions.

We refer to the appendix for further details. In Appendix \ref{sec:parameter-correlations}, we focus on the above mentioned fermionic model and analyze the relation between halo mass and halo radius which is qualitatively consistent with DM-SDR. In Appendix \ref{sec:appendix:parameter-distribution}, we provide model parameter distributions of the competing DM models.

\section{Methodology}
\label{sec:fitting}

The data used in this work is obtained from the Spitzer Photometry and Accurate Rotation Curves (\href{http://astroweb.cwru.edu/SPARC/}{SPARC}) data-set. It contains independent observations of the total velocity ($V_{\rm tot}$) and luminous mass distributions which allows to infer the bulge ($V_{\rm b}$), disk ($V_{\rm d}$), and gas ($V_{\rm g}$) velocity contributions. We extract the tangential velocities of the DM component from the data (for each galaxy of the sample) by subtracting the inferred baryonic components (as provided in the SPARC data-set) from the total velocity $V_{\rm tot}$, and statistically compare with the corresponding velocities predicted by each DM model as detailed in next. 

\subsection{Data selection}
\label{sec:data}

The SPARC data-set includes \SI{3.6}{\micro\meter} near-infrared and \SI{21}{\centi\meter} observations. The former traces the stellar mass distribution (bulge and disk) while the latter traces the atomic gas distribution and provides velocity fields from which the RCs are derived. See \citet{2016AJ....152..157L} for a complete description of the SPARC data.

The data is distributed in separated files such as \href{http://astroweb.cwru.edu/SPARC/SPARC_Lelli2016c.mrt}{Table1.mrt} (i.e. Hubble type, inclination etc.) and \href{http://astroweb.cwru.edu/SPARC/MassModels_Lelli2016c.mrt}{Table2.mrt} (i.e. RC data) and can be found at \url{http://astroweb.cwru.edu/SPARC/}.

We extract the observed circular velocity $\SYMvtot$ and the baryonic contribution $\SYMvbar$, composed of a bulge ($\SYMvbulge$), disk ($\SYMvdisk$) and gas component ($\SYMvgas$). The bulge and disk components are inferred from surface brightness observations for a given mass-to-light ratio. The baryonic component is then given by 
\begin{equation}
	\label{eqn:baryonic-sum}
	\SYMvbar^2 = \Upsilon_\mathrm{b}^{\phantom{2}} \SYMvbulge^2 + \Upsilon_\mathrm{d}^{\phantom{2}} \SYMvdisk^2 + \SYMvgas^2.
\end{equation} 
For convenience, the given velocities $\SYMvbulge$ and $\SYMvdisk$ in the SPARC data are normalized for a mass-to-light ratio of $1\,M_\odot/L_\odot$.

In this work we follow the same data selection criteria as done in \citet{2016PhRvL.117t1101M}. We choose averaged mass-to-light ratios $\Upsilon_\mathrm{b} = 0.7$ for all bulges and $\Upsilon_\mathrm{d} = 0.5$ for all disks as convenient average representatives. We exclude all galaxies with a bad quality flag ($Q=3$) and face-on galaxies with an inclination $i < \SI{30}{\degree}$. The latter is to minimize the $\sin(i)$ corrections to the observed velocities. For all measurements we require a minimum precision of \SI{10}{\percent} in velocity.

Additionally we reject all points where the baryonic velocity is greater than \SI{95}{\percent} of the observed velocity. This condition is required to avoid negative velocities for the inferred DM components. It affects mainly data points in the inner region which is dominated by baryonic matter and strongly depends on the chosen mass-to-light factors. Therefore, those inner points are less reliable. Finally, we exclude all remaining galaxies with less than 6 data points to be statistically significant.

We obtain $120$ galaxies (out of $174$) with $2396$ points (of $3355$) in total. Galaxies not fulfilling the quality criteria have such poor data that they do not allow to gain any insights. In the worst case (e.g. too few points, points on a nearly straight line, etc.) it is not possible to fit the rotation curves.

\subsection{Data fitting}
\label{LM-fitting}

We infer the circular velocity of the DM component directly from observations via 
\begin{equation}
	\label{eqn:baryonic-diff}
	\SYMvdark^2 = \SYMvtot^2 - \SYMvbar^2,
\end{equation} 
where the uncertainty in $V_{\rm DM}$ is calculated from the uncertainties in the other velocity components within linear error propagation theory. Note that only the uncertainties in the total RCs ($\Delta V_{\rm tot}$) are provided within the SPARC data-set. The uncertainty is therefore given by
\begin{equation}\label{eqn:VDM-error}
    \Delta V_{\rm DM} = \abs{\diffpart{V_{\rm DM}}{V_{\rm tot}}} \Delta V_{\rm tot}.
\end{equation} 
The inferred DM contribution of each galaxy then will be fitted by the competing DM models which are described in section \ref{sec:dark-matter-models}.

With this information, we use the Levenberg–Marquardt (LM) algorithm of least-square error minimization for each dark matter component calculated by 
\begin{equation}
    \label{eqn:method:chi-square}
	\chi^2(\vec p) = \sum \limits_{i=1}^N \qbracket{\frac{V_i - v(r_i,\vec p)}{\Delta V_{i}}}^2.
\end{equation} 
Here, $N$ is the number of data points for a given galaxy, $V_i$ is the set of inferred DM circular velocities from the data at each corresponding measured radius $r_i$, while $v(r_i,\vec p)$ accounts for the circular velocity at $r_i$ for each model parameter vector $\vec p$ (described below), and $\Delta V_{i}$ is the uncertainty in $V_i$ as given in \cref{eqn:VDM-error}.

\section{Dark matter models}
\label{sec:dark-matter-models}

Most of the DM halo models in the literature are phenomenological, i.e. motivated by the phenomenology of rotation curves either from observations or from numerical N-body simulations. In contrast, we consider a fermionic DM model based on first physical principles including statistical-mechanics and thermodynamics.

\subsection{Fermionic DM halos from MEP}
\label{sec:model:rar}

It has been proposed by several authors (see, e.g. \citealp{2020EPJP..135..290C} for an exhaustive list of references) that DM halos could be made of fermions (e.g. sterile neutrinos) in gravitational interaction. It is usually assumed that the fermions are in a statistical equilibrium state described by the Fermi-Dirac distribution function. However, the notion of statistical equilibrium for systems with long-range interactions is subtle. If the fermions are non-interacting, apart from gravitational forces, the relaxation time towards statistical equilibrium due to gravitational encounters scales as $(N/\ln N)t_D$ \citep[see e.g.][]{Binney2008} and exceeds the age of the Universe by many orders of magnitude.

For example, assuming a fermion mass $m c^2 \sim \SI{50}{\kilo\eV}$, a DM halo of mass $M\sim \SI{E11}{\Msun}$ and radius $R \sim \SI{30}{\kilo\parsec}$ contains $N\sim \num{E72}$ fermions for a dynamical time $t_D\sim 1/\sqrt{R^3/G M}\sim 100\, {\rm Myrs}$.

Therefore, on the Hubble time, the gas of fermions is essentially collisionless, being described by the Vlasov-Poisson equations. Yet, it can achieve a form of statistical equilibrium on a coarse-grained scale through a process of violent relaxation. This concept was introduced  by \cite{1967MNRAS.136..101L} in the case of collisionless stellar systems and has been exported to DM by \citet{1996ApJ...466L...1K} and \citet{2015PhRvD..92l3527C}. 

Assuming ergodicity (efficient mixing), \citet{1967MNRAS.136..101L} used a MEP and looked for the {\it most probable} equilibrium state consistent with the constraints of the collisionless dynamics. The maximization of the Lynden-Bell entropy $S$ under suitable constraints leads to a coarse-grained distribution function $\bar{f}({\vec r},{\vec v})$ similar to the Fermi-Dirac distribution function. Therefore, the process of violent relaxation may provide a justification of the Fermi-Dirac distribution function for DM halos without the need of efficient gravitational encounters.

However, when coupled to gravity, this distribution function has an infinite mass (i.e., there is no maximum entropy state), implying that either violent relaxation is incomplete or that tidal effects have to be taken into account (if the system is not isolated). The problem therefore becomes an out-of-equilibrium problem and it is necessary to develop a kinetic theory of collisionless relaxation (see e.g. \citealp{2021arXiv211213664C} for a review).

One approach is to use a Maximum Entropy Production Principle (MEPP) and argue that the most probable evolution of the system on the coarse-grained scale is the one that maximizes the rate of Lynden-Bell entropy $\dot S$ under the constraints of the collisionless dynamics \citep{1996ApJ...471..385C}. This leads to a generalized Fokker-Planck equation having the form of a fermionic Kramers equation 
\begin{equation}
    \label{kramers}
     \frac{\partial \bar f}{\partial t} + \vec v \cdot \frac{\partial \bar f}{\partial \vec r} - \nabla\Phi \cdot \frac{\partial \bar f}{\partial \vec v} = \frac{\partial \vec J}{\partial \vec v},
\end{equation}
where $\vec J=D[\partial \bar f/\partial \vec v + (mc^2/kT) \bar f (1-f/\eta_0) \vec v]$ is a diffusion current pushing the system towards statistical equilibrium, $D$ the diffusion coefficient, $T \equiv T(t)$ is the temperature evolving in time so as to conserve the total energy \citep{1998MNRAS.300..981C}, $k$ is the Boltzmann constant, $c$ is the speed of light, and $m$ is the DM fermion mass. However, this approach is heuristic and does not determine the expression of the diffusion coefficient.

An alternative, more systematic, approach is to develop a quasilinear theory of ``gentle'' collisionless relaxation \citep{1970PhRvL..25.1155K,1980Ap&SS..72..293S,1998MNRAS.300..981C,2004PhyA..332...89C} leading to a fermionic Landau equation of the form 
\begin{multline}
    \label{landau}
    \frac{\partial \bar f}{\partial t} + \vec v \cdot \frac{\partial \bar f}{\partial \vec r} - \nabla\Phi \cdot \frac{\partial \bar f}{\partial \vec v}
    = \frac{8\pi G^2m^8\epsilon_r^3\epsilon_v^3\ln\Lambda}{h^6}\frac{\partial}{\partial v_i} \int {\rm d}\vec v'\\
    \times \frac{u^2\delta_{ij} - u_i u_j}{u^3}\left\lbrace \bar f' \left (1 - \bar f'\right ) \frac{\partial \bar f }{\partial v_j} - \bar f \left(1-\bar f\right) \frac{\partial \bar f'}{\partial
    v'_j}\right\rbrace,
\end{multline} 
where $\bar{f}' \equiv \bar{f}({\vec r},{\vec v}', t)$, $\ln\Lambda=\ln(R/\epsilon_r)$ is the Coulomb logarithm, $R$ is the typical size of the system, $\vec u = \vec v' - \vec v$ is the relative velocity between the ``macro-particles'' of mass $m_{\rm eff}\sim 2m^4 \epsilon_r^3\epsilon_v^3/h^3 \gg m$ (see also below), and $\epsilon_r$, $\epsilon_v$ are the correlation lengths in position and velocity respectively. One can make a connection between the above two kinetic equations by using a form of thermal bath approximation, i.e., by replacing $\bar f'$ in \cref{landau} by its equilibrium (Fermi-Dirac) expression. This substitution transforms an integro-differential (Landau) equation into a differential (Kramers) equation. In this manner one can compute the diffusion coefficient explicitly \citep{1998MNRAS.300..981C}.

The timescale of violent relaxation is a few $10-100$ dynamical times ($t_D$), which is shorter than the Hubble time $t_H$. This is confirmed by the kinetic theory of violent relaxation that predicts a collisionless relaxation time $t_R^{\rm non-coll.}\sim (M/m_{\rm eff})t_D$ which is much shorter than the collisional relaxation time $t_R^{\rm coll.}\sim (M/m)t_D$ because $m_{\rm eff}\gg m$ (see formula in the above paragraph).

Indeed, the relaxation of the coarse-grained DF $\bar{f}({\vec r},{\vec v},t)$ towards the Lynden-Bell distribution (of Fermi-Dirac type, see Eq. \ref{fcDF} below) on a few dynamical times can be interpreted in terms of ``collisions'' between  ``macro-particles'' or ``clumps'' (i.e. correlated regions) with a large effective mass $m_{\rm
eff}$ \citep{1970PhRvL..25.1155K}. These macro-particles considerably accelerate the relaxation of the system (as compared to ordinary gravitational encounters between particles of mass $m$) by increasing the diffusion coefficient $D$ in \cref{kramers}. 

Processes of incomplete relaxation could be taken into account by generalizing the kinetic approach so that the diffusion coefficient rapidly falls off to zero in space and time, thereby leading to a sort of kinetic blocking. Alternatively, if the system is submitted to tidal interactions from neighboring systems one can look for a stationary solution of \cref{kramers} which accounts for the depletion of the distribution function above an escape energy.

For classical systems evolving through two-body gravitational encounters like globular clusters, this procedure leads to the King model \citep{1962AJ.....67..471K}. For fermionic DM halos, one obtains the fermionic King model \citep{1983A&A...119...35R,1998MNRAS.300..981C} 
\begin{equation}
    \bar{f}(r,\epsilon\leq\epsilon_c) = \frac{1-e^{[\epsilon-\epsilon_c(r)]/kT(r)}}{e^{[\epsilon-\mu(r)]/kT(r)}+1}, \qquad \bar{f}(r,\epsilon>\epsilon_c)=0\, ,
    \label{fcDF}
\end{equation} 
which has been written in the case of general relativistic fermionic systems for the sake of generality \citep{2018PDU....21...82A,2022IJMPD..3130002A}. Here, $\epsilon=\sqrt{p^2c^2 + m^2 c^4} - mc^2$ is the particle kinetic energy, $\mu(r)$ is the chemical potential (with the particle rest-energy subtracted off), $\epsilon_c(r)$ is the escape energy (with the particle rest-energy subtracted off), and $T(r)$ is the effective temperature. The corresponding set of three dimensionless parameters (for fixed $m$) are defined by the temperature, degeneracy and cutoff parameters, $\beta(r)=k T(r)/(m c^2)$, $\theta(r)=\mu(r)/[k T(r)]$ and $W(r)=\epsilon_c(r)/[k T(r)]$, respectively (a subscript $0$ is used when the parameters are evaluated at the center of the configuration).

This distribution function takes into account the Pauli exclusion principle as well as tidal effects, and can lead to a relevant model of fermionic DM halos usually referred as the RAR model, which has been successfully contrasted against galaxy observables \citep{2018PDU....21...82A,2019PDU....24..278A,2020A&A...641A..34B,2021MNRAS.505L..64B,2022MNRAS.511L..35A}.

The full family of density $\rho(r)$ and pressure $P(r)$ profiles within this model can be directly obtained as the corresponding integrals of $\bar{f}(p)$ over momentum space (bounded from above by $\epsilon \leq \epsilon_c(r)$) as detailed in \citet{2018PDU....21...82A}. This leads to a four-parametric fermionic equation of state depending on ($\beta_0,\theta_0,W_0,m$) according to the parameters in \cref{fcDF}. Once with the fermionic distribution function at equilibrium as obtained from the MEP explained above, we make use of the fact that a relaxed system of fermions under self-gravity does admit a perfect fluid approximation \citep{1969PhRv..187.1767R}. Thus, we use the stress-energy tensor of a perfect fluid in a spherically symmetric metric, $g_{\mu\nu}=\diag(\e^{2\nu(r)},-\e^{2\lambda(r)},-r^2,-r^2\sin^2\vartheta)$ with $\nu(r)$, $\lambda(r)$ being the temporal and spatial metric functions and $\vartheta$ the azimutal angle. Such configuration leads to hydrostatic equilibrium equations of self-gravitating fermions. The local $T(r)$, $\mu(r)$ and $W(r)$ fulfill the Tolman, Klein and particle's energy conservation relations, respectively (see \citealp{2018PDU....21...82A} for details). Further, $\nu_0$ is here constrained by the Schwarzschild condition $g_{00}g_{11} = -1$ at the surface where the halo pressure (and density) falls to zero.

\begin{figure}
	\centering%
	\includegraphics[width=\hsize]{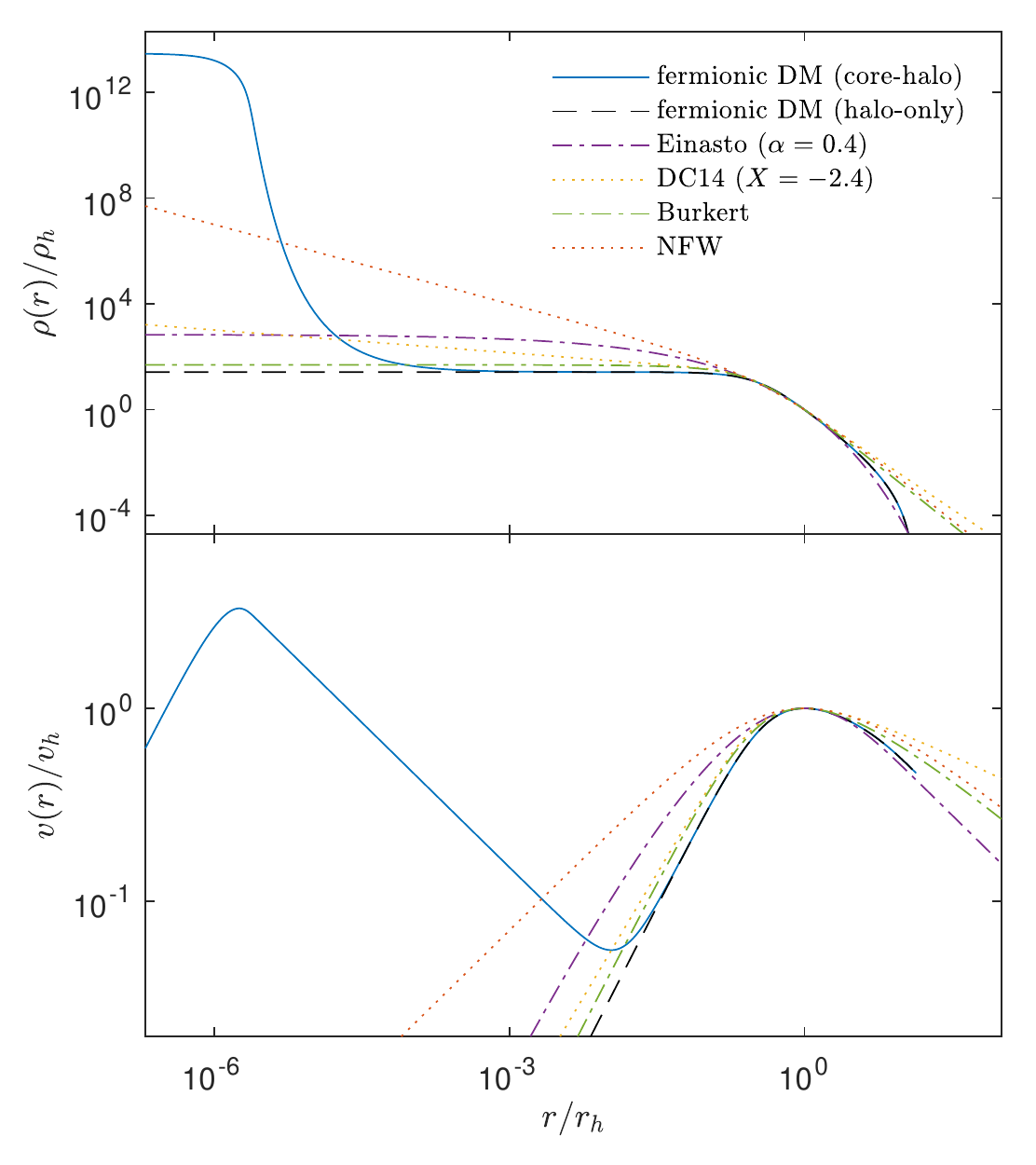}
	\caption{Illustrations of DM models: density profile (top) and rotation curve (bottom). Shown are typical configuration parameters as obtained from this work and detailed in section \ref{sec:appendix:parameter-distribution}. The core-halo solution of the fermionic DM model is illustrated by the configuration parameters $\beta_0 = \num{E-7}$, $\theta_0 = 30$ and $W_0 = 60$, in combination with a corresponding halo-only solution. For comparison, other DM models (NEW, DC14, Burkert, Einasto) are added. All profiles are normalized with respect to the halo radius $r_h$, defined at the halo velocity maximum, with $\rho_h = r(r_h)$ and $v_h = v(r_h)$.}%
	\label{fig:profile-illustration-mep}%
\end{figure}

An illustration of a core-halo ($\theta_0 > 10$) and a corresponding halo-only ($\theta_0 \ll -1$) solution are shown in \cref{fig:profile-illustration-mep} (we refer to section \ref{sec:morph} and to the cited works above to get a better understanding of the rich morphology of the fermionic DM model).

\def\ROOTPATH{figure/ProfileIllustrationsAll}

Mass distributions of that fermionic model are well characterized by the cutoff difference $W(r_p) - W(r_s)$. Since $W(r)$ is defined to be zero at the surface, i.e. $W(r_s) = 0$ with $r_s$ being the surface radius where the density drops to zero, we need to focus only on the plateau cutoff $W_p = W(r_p)$ with $r_p$ being the plateau radius defined at the first minimum in the rotation curve. 

Other important quantities of the \textit{core-halo} family of fermionic DM mass solutions are the core mass $M_c = M(r_c)$ with $r_c$ being the core radius defined at the first maximum in the rotation curve, the halo mass $M_h = M(r_h)$ with $r_h$ being the halo radius defined at the second maximum in the rotation curve, and the total mass $M_s = M(r_s)$ given at the surface radius $r_s$.

The density profiles in the fermionic model can develop a rich morphological behaviour: while the halo region is King-like (i.e. from polytropic-like with $W_p \ll 1$ to power law-like with $W_p \gtrsim 10$, see \ref{sec:morph}), the inner region can either develop a \textit{dense core} at the center of such a halo (i.e. for large central degeneracy $\theta_0 > 10$), or not (i.e. $\theta_0 \ll -1$ in the dilute regime). Both kind of family profiles are thermodynamically and dynamically stable as well as long lived in a cosmological framework, as recently demonstrated in \citet{2021MNRAS.502.4227A} for typical galaxies with total masses of the order $\sim \SI{5E10}{\Msun}$.

Remarkably, for \textit{core} - \textit{halo} RAR solutions with fermion masses of $m c^2\approx \SI{50}{\kilo\eV}$, the degenerate and compact DM cores may work as an alternative to the BH paradigm at the center of non-active galaxies \citep{2018PDU....21...82A,2019PDU....24..278A,2020A&A...641A..34B,2021MNRAS.505L..64B,2022MNRAS.511L..35A}. Furthermore, their eventual gravitational core-collapse in larger galaxies may offer a novel supermassive BH formation mechanism from DM \citep{2021MNRAS.502.4227A}.

\subsection{Other DM halo models}
\label{sec:model:dc14-nfw}

From a phenomenological viewpoint, a DM halo density profile is described by three characteristics: the inner halo, the outer halo and the transition in between. Such density profiles are usually described by the ($\alpha, \beta, \gamma$)-model \citep{1990ApJ...356..359H} 
\begin{equation}
    \label{eqn:hernquist}
	\frac{\rho(r)}{\rho_\mathrm{N}} = \qbracket{\frac{r}{R_\mathrm{N}}}^{-\gamma}\qbracket{1 + \qbracket{\frac{r}{R_\mathrm{N}}}^\alpha}^{-\frac{\beta - \gamma}{\alpha}},
\end{equation} 
where $\alpha$ describes the transition, $\beta$ the outer slope and $\gamma$ the inner slope. Following Newtonian dynamics --- that is fully sufficient on halo scales --- then the velocity is given by \begin{equation}
    \label{eqn:circular-velocity}
    \frac{v^2(r)}{\sigma_\mathrm{N}^2} = \frac{R_{\rm N}}{r}\frac{M(r)}{M_{\rm N}}
\end{equation} and the enclosed mass by \begin{equation}
    \label{eqn:enclosed-mass}
     \frac{M(r)}{M_{\rm N}} = \int \limits_0^r \qbracket{\frac{r}{R_{\rm N}}}^2 \frac{\rho(r)}{\rho_\mathrm{N}} \frac{\d r}{R_{\rm N}},
\end{equation} For the following DM halo models we will use $R_\mathrm{N}$, $\rho_\mathrm{N}$, $\sigma_\mathrm{N}^2 = G M_\mathrm{N}/R_\mathrm{N}$ and $M_\mathrm{N} = 4\pi \rho_\mathrm{N} R_\mathrm{N}^3$ as scaling factors for length, density, velocity and mass, respectively.

In \cref{fig:profile-illustration-mep} we illustrate the typical morphology of common DM halo models used here for the SPARC galaxies. For a better comparison the plots are normalized with respect to the halo located at the velocity maximum on halo scales.

Based on the general ($\alpha, \beta, \gamma$)-model, \citet{2014MNRAS.441.2986D} modelled CDM halos including baryonic feedback mechanisms in galaxy formation. They found that the three parameters ($\alpha, \beta, \gamma$) are related through 
\begin{align}
    \label{eqn:dc14:alpha}
	\alpha &= 2.94 - \log_{10}\qbracket{(10^{X + 2.33})^{-1.08} + (10^{X+2.33})^{2.29}},\\
	\label{eqn:dc14:beta}
    \beta &= 4.23 + 1.34 X + 0.26 X^2,\\
    \label{eqn:dc14:gamma}
    \gamma &= -0.06 + \log_{10}\qbracket{(10^{X + 2.56})^{-0.68} + 10^{X+2.56}},
\end{align} 
where $X = \log_{10}(M_*/M_\mathrm{halo})$ describes the stellar-to-dark matter ratio. For the circular velocity \cref{eqn:circular-velocity} the enclosed mass \cref{eqn:enclosed-mass} is given by a hypergeometric function
\begin{equation}
	\frac{M(r)}{M_\mathrm{N}} = \frac{1}{3-\gamma} \qbracket{\frac{r}{R_\mathrm{N}}}^{3-\gamma} \pFq{2}{1}(p_1,p_2;\,q_1;\,-[r/R_\mathrm{N}]^\alpha),
\end{equation} 
with $p_1 = (3-\gamma)/\alpha$, $p_2 = (\beta-\gamma)/\alpha$ and $q_1 = 1 + (3 - \gamma)/\alpha$. In the following, we refer this model as DC14.

Alternatively, for $\alpha = 1$, $\beta = 3$ and $\gamma = 1$ the ($\alpha, \beta, \gamma$)-model reduces to the NFW model \citep{1996ApJ...462..563N,1997ApJ...490..493N} as obtained from early DM-only N-body simulations. This DM model develops cuspy halos of the following type
\begin{equation}
	\frac{\rho(r)}{\rho_\mathrm{N}} = \qbracket{\frac{r}{R_\mathrm{N}}}^{-1}\qbracket{1 + \frac{r}{R_\mathrm{N}}}^{-2},
\end{equation} 
with the circular velocity \begin{equation}
	\frac{v^2(r)}{\sigma_\mathrm{N}^2} = \frac{\ln(1 + r/R_\mathrm{N})}{r/R_\mathrm{N}} - \frac{1}{1 + r/R_\mathrm{N}}.
\end{equation} 

In contrast to NFW, \citet{1995ApJ...447L..25B} proposed a DM density profile with a cored halo of the following type
\begin{equation}
    \frac{\rho(r)}{\rho_\mathrm{N}} = \qbracket{1 + \frac{r}{R_\mathrm{N}}}^{-1}\qbracket{1 + \bracket{\frac{r}{R_\mathrm{N}}}^{2}}^{-1}.
\end{equation} 
For the circular velocity \cref{eqn:circular-velocity} the enclosed mass \cref{eqn:enclosed-mass} is given by
\begin{multline}
     \frac{M(r)}{M_{\rm N}} = \frac14 \ln(1 + [r/R_{\rm N}]^2) + \frac12 \ln(1 + r/R_{\rm N})\\
     - \frac12 \arctan(r/R_{\rm N}).
\end{multline} 
With $M_0 \approx M(R_N)$ being the mass scale originally interpreted as the core mass of the halo \citep[e.g.][]{2000ApJ...537L...9S} we obtain the relation $M_N = 8 M_0$. Further, the density scale $\rho_N$ describes the central density $\rho_0$ and the length scale $R_{\rm N}$ can be identified with the Burkert radius $r_{\rm B}$ fulfilling the condition $\rho(r_{\rm B}) = \rho_0/4$.

Another interesting and successful candidate is the Einasto model \citep{1989A&A...223...89E}, a purely empirical fitting function with no commonly recognized physical basis \citep{2006AJ....132.2685M}. The DM halo density profiles of that model are of the following type, given in a normalized form,
\begin{equation}
    \frac{\rho(r)}{\rho_\mathrm{N}} = \e^{-[r/R_{\rm N}]^{\kappa}}.
\end{equation} 
The exponent $\kappa$ describes the shape of the density profile. The circular velocity and the enclosed mass are given by \cref{eqn:circular-velocity,eqn:enclosed-mass}. This model develops mass distributions with a finite mass $M_{\rm tot}/M_{\rm N} = \Gamma(3/\kappa)/\kappa$ for $r\to\infty$ (see also \citealp{2012A&A...540A..70R}). The typical $\kappa$ values obtained in this work for the SPARC data-set, as well as the comparison with the same values coming from N-body simulations (either with or without baryonic effects), are given in subsections \ref{boundaryC}, \ref{sec:baryonic-effect} and \ref{sec:result:gof}, and Fig. \ref{fig:parameter-distribution:einasto}.

\subsection{Fitting priors and Monte-Carlo approach}
\label{boundaryC}

For the fermionic DM model we fix the particle mass $m$ and therefore reduce the number of free parameters by one, e.g. $\vec p = (\beta_0, \theta_0, W_0)$. A particle mass of $mc^2 = \SI{50}{\kilo\eV}$ is well motivated by the promising results obtained in \citet{2018PDU....21...82A,2020A&A...641A..34B,2021MNRAS.505L..64B,2022MNRAS.511L..35A}, where the fermionic core-halo DM profile was able to explain both the S-stars orbits around SgrA*, and the Milky Way rotation curve. In \citet{2019PDU....24..278A,2021MNRAS.502.4227A}, in particular, and for the same particle mass, the fermionic core-halo profiles were successfully applied to other galaxy types from dwarf to larger galaxy types, providing a possible explanation for the nature of the intermediate mass BHs, as well as a possible mechanism for SMBH formation in active galaxies. Moreover, as demonstrated in \citet{2018PDU....21...82A} there exists a particle mass range between $\sim$ 50 and $\sim$ 350 keV where the compacity of the central core can increase (all the way to its critical value of collapse) while maintaining the same DM halo-shape. Therefore, regarding the SPARC RC fitting, as well as all the scaling relations on halos-scales, our conclusions are not biased by the choice of the particle mass in the above range.

For the fermionic model we consider solutions which are either in the dilute regime ($\theta_0 \ll -1$, i.e. are King-like), or which have developed a degenerate core (i.e. $\theta_0 > 10$) at the center of such a halo. Fermionic solutions within only these two families have been shown to be thermodynamically and dynamically stable when applied to galaxies \citep{2021MNRAS.502.4227A}.

The NFW and the Burkert models are described by two free scaling parameters, e.g. $\vec p = (R_{\rm N},\rho_{\rm N})$. The DC14 model with e.g. $\vec p = (X, R_{\rm N}, \rho_{\rm N})$ and the Einasto model with e.g. $\vec p = (\kappa, R_{\rm N}, \rho_{\rm N})$ are described by three free parameters. Compared to NFW and Burkert, both (Einasto and DC14) have an additional parameter which affects the sharpness of the transition from the inner to the outer halo.

In order to find the best-fits we use the LM algorithm (see section \ref{LM-fitting}) with well chosen initial values (i.e. priors) reflecting astrophysical (realistic) scenarios. Because the LM algorithm finds only local minima, we choose the parameter sets randomly within a range and follow a Monte-Carlo approach. For the fermionic \textit{core}-\textit{halo} solutions, we choose $\beta_0 \in [10^{-8},10^{-5}]$, $\theta_0\in [25,45]$ and $W_0\in[40,200]$ which correspond to a conservatively wide range of parameters according to \citet{2019PDU....24..278A}. For the fermionic \textit{diluted} solutions, we cover the same range for $\beta_0$ and $W_0$, but with $\theta_0\equiv\theta_p \in [-40,-20]$. For the other DM models, the initial scaling factors are chosen from $R_{\rm N}\in[10^{1}, 10^{4}]\si{\parsec}$ and $\rho_{\rm N} \in [10^{-4}, 10^{-1}]\si{\Msun\per\parsec^3}$. According to \citet{2017MNRAS.466.1648K} the additional parameter of the DC14 model is chosen from $X\in[-3.75,-1.3]$. For the Einasto model, the exponent is chosen from $\kappa\in[0.1, 10]$. This relatively large window has been chosen in order (i) to account for the broad diversity of rotation curves covered by the SPARC data, and (ii) not to be limited by any fixed value (e.g. as typically obtained by CDM-only simulations) since this is an independent analysis to that of N-body simulations, and may also account for other effects such as baryonic feedback (see also the discussion in next subsection).

\subsection{Baryonic feedback}
\label{sec:baryonic-effect}

\begin{figure*}
	\centering%
	\includegraphics[width=\hsize]{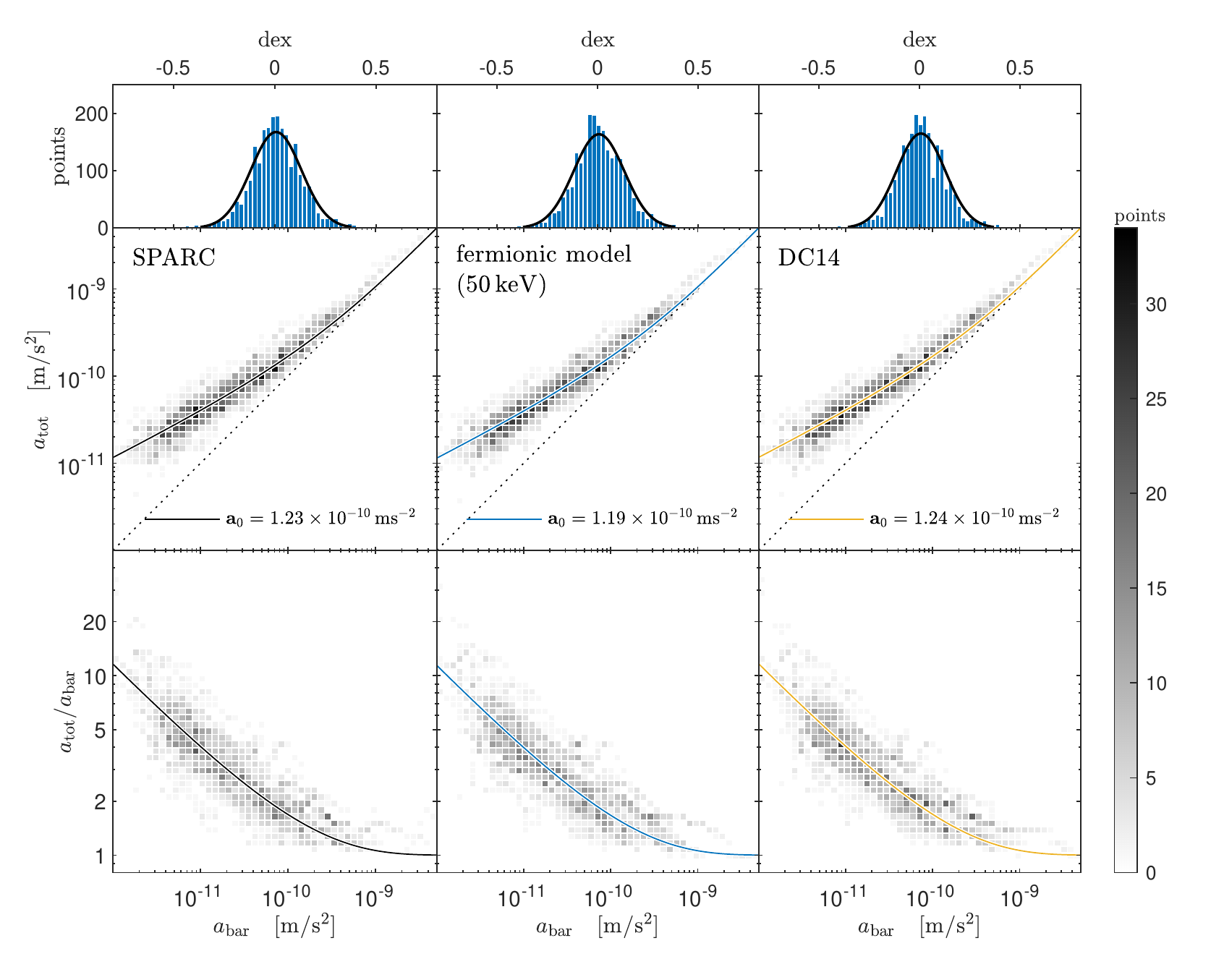}
	\caption{Radial acceleration relation (top) and mass discrepancy acceleration relation (bottom) for SPARC data and competing DM halo models. Each plot is divided in 50x50 equal bins. The baryonic centripetal acceleration $\SYMabar$ is inferred from luminosity observables while the total acceleration $\SYMatot$ is inferred independently from velocity fields. For DM halo models the total acceleration is composed of the predicted dark and inferred baryonic components, i.e. $\SYMatot = \SYMabar + \SYMaDM$. The corresponding solid curves are the best fits characterized by a specific $\mathfrak{a}_0$. The histogram plots (upper row) show a Gaussian distribution of $\log_{10}(\SYMatot/\mathfrak{a}_0)$. The grayscale legend shows the number of points per bin (of a total 2396 for the 120 SPARC-galaxies used).}%
	\label{fig:acceleration:grid}%
\end{figure*}

The DM in galaxies evolves together with baryons, thus it is expected some degree of baryonic feedback on scales where baryons dominate. This also seems to be the case within the SPARC galaxies here considered, where almost half of the points in upper-left panel of Fig. \ref{fig:acceleration:grid} 
roughly fulfill $\SYMatot < 2 \SYMabar$. One of the main baryonic effects onto the total gravitational potential is thought to happen due to a sustained process of stellar bursts, which drives baryonic material from inner-halo regions while ending in a reduction of DM densities at those halo scales (see e.g. \citealp{2012MNRAS.422.1231G,2016MNRAS.459.2573R}). However, the quantification of such baryonic effects have always been calibrated and applied to cuspy CDM halos. Only recently it was shown that baryonic feedback is DM-model dependent.

That is, in WDM cosmologies such effects are typically diminished with respect to CDM \citep{2019MNRAS.483.4086B}. The main reasons behind this attenuation are that in WDM cosmologies DM halos form later, are less centrally dense on inner-halo scales, and therefore contain galaxies that are less massive with less baryon content than the CDM counterparts. Since the halos of the Burkert, Einasto and fermionic DM model correspond to WDM cosmologies, it is expected that baryonic effects are milder in those cases. However, a thorough quantification of this feedback has only been worked out for Einasto profiles \citep{2019MNRAS.483.4086B} and still remains to be solved in the case of fermionic halos before taking any conclusion, though out of the scope of this work.

In any case, and regardless of a potential feedback of the baryons onto the fermionic DM profile, it is important to emphasize a key result of this work: the flatness in the inner-halo slope of MEP profiles is due to a (quasi) thermodynamic equilibrium reached by the DM particles, thus involving a different physical principle than the one explained above for the baryons.

Finally, two DM model considered here --- DC14 \citep{2014MNRAS.441.2986D} and Einasto \citep{2019MNRAS.483.4086B} --- (effectively) account for baryonic feedback. That is, any baryonic feedback expected to arise for SPARC galaxies should be reflected in the best-fit parameters of the DM profiles with its consequent universal shape reflecting such effects. In section \ref{sec:result:gof} we compare our statistical results regarding baryonic effects in the DM model free-parameters with the ones reported in the literature.



\def\ROOTPATH{figure/AccelerationGrid}
\begin{figure*}
	\figurenum{\arabic{figure}}
	\centering%
	\includegraphics[width=\hsize]{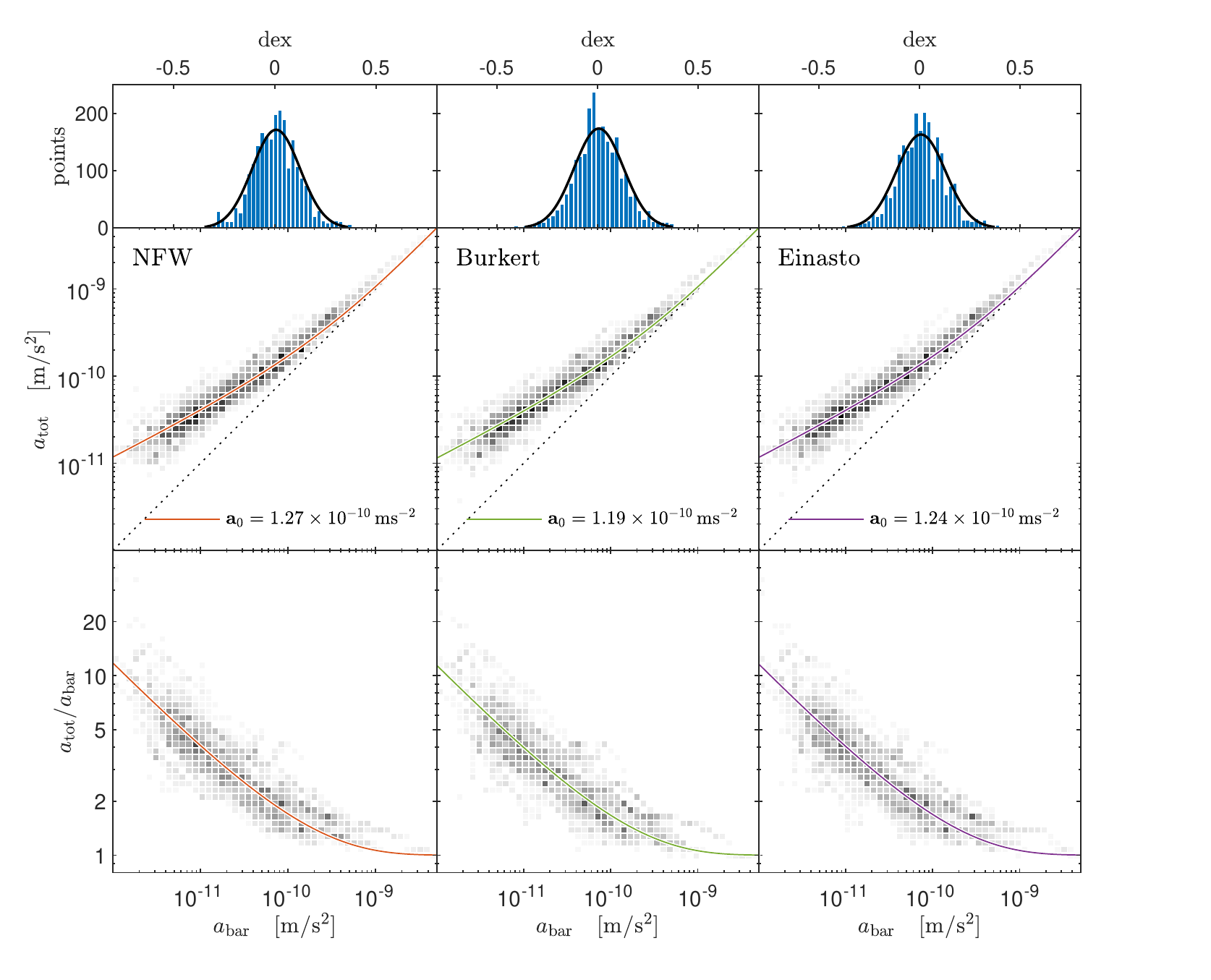}
	\caption{continue}
	\label{fig:acceleration:grid-b}%
\end{figure*}

\section{Results}
\label{sec:results}

After an insightful analysis of the parameter distribution of each DM model which best fits the SPARC RCs (given in appendix \ref{sec:appendix:parameter-distribution}), we compare between them following two complementary approaches.

First, we consider the entire galaxy sample and extract the radial acceleration information for the total and baryonic components at each galactocentric radii, and put them all together as in \cref{fig:acceleration:grid} (SPARC-window). In this way, we reduce any characteristics of individual galaxies into an overall (\textit{global}) picture, i.e. the Radial Acceleration Relation \citep{2016PhRvL.117t1101M} and Mass Discrepancy Acceleration Relation \citep{2004ApJ...609..652M,2014Galax...2..601M}, respectively.

In the second approach, we consider each \textit{individual} galaxy and perform a goodness of DM-model analysis, showing how well the inferred DM RCs are fitted by these models.

\subsection{Acceleration relations}
\label{sec:result:ac}

The rotation curve of each component (e.g. bulge, disk, gas) traces its centripetal acceleration $\SYMacc = v^2/r$, giving access to independent acceleration measurements. The Radial Acceleration Relation compares the radial acceleration due to the total mass ($\SYMatot$) with that due to the baryonic mass ($\SYMabar$). It is empirically described by \cite{2016PhRvL.117t1101M}
\begin{equation}
	\label{eq:mcgaugh-fit}
	\SYMatot = \frac{\SYMabar}{1 - \e^{-\sqrt{\SYMabar/{\SYMafrak}}}},
\end{equation}
where $\SYMafrak$ is the only adjustable parameter. In the low acceleration regime ($\SYMabar \ll \SYMafrak$), where DM dominates, it clearly shows a deviation from a linear correlation (see top panels of \cref{fig:acceleration:grid}). While in the high acceleration regime ($\SYMabar \gg \SYMafrak$), dominated by baryonic matter, the linear relation is recovered.

This relation is not limited to disk galaxies but also holds for other galaxy types (e.g. ellipticals, lenticulars, dwarfs spheroidals and even low-surface-brightness galaxies), what makes it a true universal law among morphology classification \citep{2016PhRvL.117t1101M,2017ApJ...836..152L,2019ApJ...873..106D}.


For the SPARC galaxies the different accelerations $\SYMabar = V_{\rm bar}^2/r$ and $\SYMatot = V_{\rm tot}^2/r$ are inferred from the SPARC data. See section \ref{sec:data} for details how the circular velocities and radii are obtained. For the competing DM models $\SYMabar = V_{\rm bar}^2/r$ is equally inferred from the SPARC data while $\SYMatot = [V_{\rm bar}^2 + v_{\rm DM}(r)^2]/r$ is inferred from the total mass distribution, composed of the observed baryonic component (taken from SPARC data) and the best-fitted circular velocities of the dark matter component $v_{\rm DM}(r)$ for each DM model. See sections \ref{sec:fitting} and \ref{boundaryC} for details how best-fits are obtained.

For the SPARC data as well as for each DM model, we applied then a least-square fitting to obtain $\SYMafrak$. The result obtained here from the SPARC data only (i.e. without assuming any specific DM model) is fully consistent (within errors) with $\SYMafrak \approx \SI{1.2E-10}{\metre\per\second^2}$ as obtained originally in \citet{2016PhRvL.117t1101M}, validating our procedure. Each DM model is then characterized by a specific best-fit $\SYMafrak$. The corresponding curves are plotted as solid lines in \cref{fig:acceleration:grid}. In all cases, we obtain values close to the one obtained in \citet{2016PhRvL.117t1101M}. This allows to conclude that all competing DM models are able to reproduce the Radial Acceleration Relation. Moreover, they reproduce it equally good without a clear statistically preferred model.

A closely related relation is the MDAR relation between the baryonic and total mass components, defined by $D = M_{\rm tot}/M_{\rm bar}$ with $M_{\rm bar}$ being the total baryonic mass and $M_{\rm tot}$ the total galaxy mass accounting for baryonic and dark matter. For $a = \diff{\Phi}{r}$ and $\Phi(r)$ being the gravitational potential of a spherically symmetry mass distribution the mass discrepancy can be equivalently written as $D = \SYMatot/\SYMabar$. The results are illustrated in the bottom panels of \cref{fig:acceleration:grid}.

According to some authors, the above two acceleration relations do not imply the need of any new physics and may be explained within the $\Lambda$CDM framework \citep{2016MNRAS.456L.127D,2017MNRAS.466.1648K,2016arXiv161208857S,2018FoPh...48.1517S}. For smaller disk and LSB galaxies (extending the original SPARC sample), \citet{2019ApJ...873..106D} found that \cref{eq:mcgaugh-fit} is a limiting case of a more complex relation with the need of adding one extra galaxy parameter. In addition, based on modern cosmological simulations, \citet{2017ApJ...835L..17K} predict even a redshift dependency of the acceleration parameter $\SYMafrak$, emphasizing that the correlation is universal only regarding the morphological classification.

Additionally to the acceleration relations mentioned above (which account for the entire accelerations distributions among all galaxies and at different radii), the next strategy is to focus on the DM components of each galaxy and gather best-fits of the inferred rotation curves allowing for another (related) quantitative comparison of the DM models.

\subsection{Goodness of model}
\label{sec:result:gof}

The SPARC galaxies show different characteristics in their rotation curve such as a nearly flat curve through the entire galaxy data; a rising trend in the inner halo followed by a single maximum; or multiple extrema in the form of oscillations. See \cref{fig:benchmark:total-rotation-curves} for three typical examples within the SPARC data-set. Some galaxies show just a rising trend implying that the rotation curves are incomplete, likely due to the faintness and/or lack of data for outermost halo stars. Of interest is therefore a quantitative description about the goodness of a DM halo model fitting the entire galaxy sample (120 galaxies).

The goodness of a fit for a single galaxy is well described by the $\chi^2$ value, see \cref{eqn:method:chi-square}. When competing models with different number of parameters are compared it is appropriate to consider the reduced $\chi^2$ defined as $\chi_r^2 = \chi^2/d$ with the degree of freedom $d = N-p$, $N$ being the number of observables (for a single galaxy) and $p$ the number of parameters (of the considered model).

The question now arises how to compare the competing models for a population of galaxies. In order to find the goodness of a model which is robust against outliers, we ask \textit{how many} fitted galaxies have a (reduced) $\chi^2$ \textit{lower} than a given one. It turns out that the population curve resembling a cumulative distribution function (CDF) follows nearly a log-normal distribution. We use the mean value, labelled as $\hat \chi^2_r$, as the criteria to described the goodness of a model for fitting a population of galaxies within the SPARC data-set. A parameter analysis of the best-fit solutions for each DM halo model is detailed in appendix \ref{sec:appendix:parameter-distribution}.

\def\ROOTPATH{figure/GoodnessAll}\begin{figure}%
	\centering%
	\includegraphics[width=\hsize]{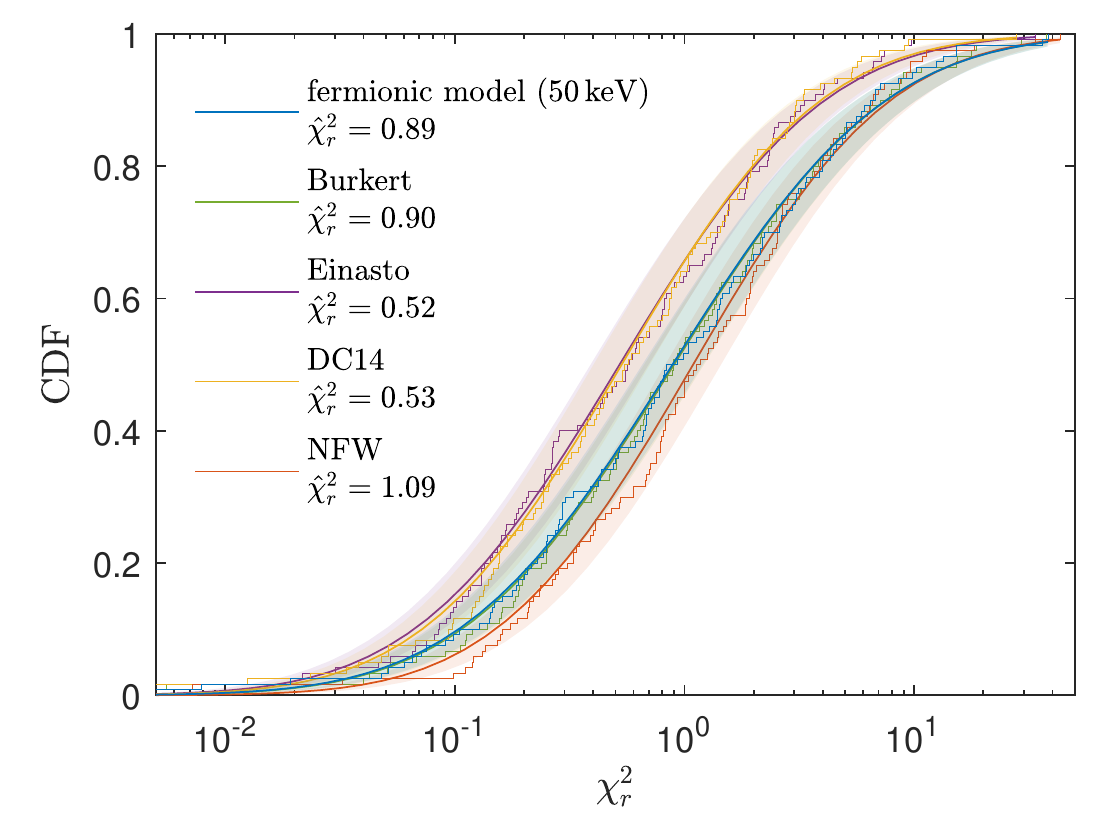}%
	\caption{Goodness of model analysis for the entire sample (120 galaxies). We count the population of fitted galaxies having a reduced $\chi^2$ smaller than a given one. The normalized population (step-like) follows nearly a log-normal distribution (solid) characterized by the mean value $\hat \chi^2_r$. The shaded regions span the $\SI{95}{\percent}$ confidence interval of the best-fitted $\hat \chi^2_r$.}%
	\label{fig:goodness:all}
\end{figure}
\def\ROOTPATH{figure/GoodnessWithCutoff}\begin{figure}%
	\centering%
	\includegraphics[width=\hsize]{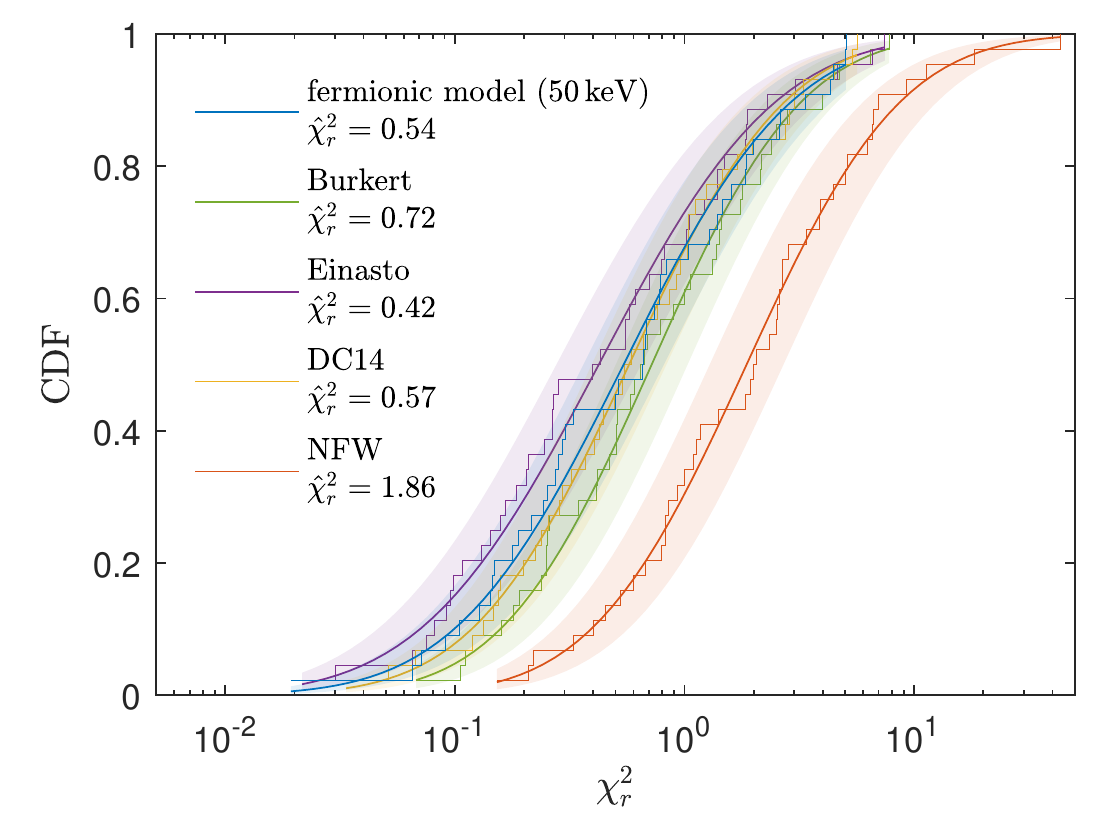}%
	\caption{Goodness of model analysis for galaxies showing one clear maximum in the outer halo RC, where DM typically dominates. This condition is fulfilled by 44 galaxies (see e.g. UGC05986 in \cref{fig:benchmark:total-rotation-curves}). We count the population of fitted galaxies having a reduced $\chi^2$ smaller than a given one. The normalized population (step-like) follows nearly a log-normal distribution (solid) characterized by the mean value $\hat \chi^2_r$. The shaded regions span the $\SI{95}{\percent}$ confidence interval of the best-fitted $\hat \chi^2_r$.}%
	\label{fig:goodness:with-cutoff}
\end{figure}

The goodness analysis for the entire SPARC sample (120 galaxies) is shown in \cref{fig:goodness:all} and can be summarized as: The NFW model ($p=2$) is statistically disfavoured with respect to the other DM halo models; the Einasto model ($p=3$) and the DC14 model ($p=3$), on the other hand, are statistically favoured with similarly good results; the Burkert model ($p=2$) and the fermionic model ($p=3$) produce comparable results but are somewhat statistically less favoured. 

We recall that the DM models are suited for DM halo rotation curves with only one maximum. Therefore, we restrict further the SPARC sample to galaxies (44) showing one clear maximum in their rotation curve (see also section \ref{sec:fermionic-halos:diversity} for details about the diversity of SPARC galaxies). This conditions is equivalently reflected in the fermionic scenario where data supports for a significant escape of DM particles, that is for energy-cutoff values at plateau of $W_p < 10$ (see appendix \ref{sec:appendix:parameter-distribution} for further details). For that sub-sample the picture changes considerably, see \cref{fig:goodness:with-cutoff}, and can be summarized as: The NFW model is statistically even more disfavoured while all other models (including fermionic DM) become more comparable --- with a little tendency for Einasto and against Burkert.

We remind that the DC14 model is based on the analysis of hydro-dynamically simulated galaxies including complex baryonic feedback processes \citep{2014MNRAS.441.2986D}. The good performance of DC14 may indicate the importance of baryonic feedback in galaxy formation. Indeed, our results regarding DC14 are in line with the literature as well since, as can be seen from \cref{fig:parameter-distribution:dc14} (bottom panel), the bulk of our ($\alpha,\beta,\gamma$) parameters for the SPARC data-set lies within the windows (0,2.6); (2.3,4); (0,2) in rough agreement with \citet{2014MNRAS.441.2986D}. Interestingly, we obtain a mean for the stellar to DM mass ratio $X = \log_{10}(M_*/M_\mathrm{halo})$ of $-2.4$, which is within the $X$ values ($-2.5,-2.3$) where DM cusps are most effectively flattened due to baryonic effects as reported in \citet{2014MNRAS.441.2986D}. Besides, our results are in good agreement with those reported in \citet{2013ApJ...770...57B} from abundance matching, since they obtain X values between (-2.7, -1.7) for halo masses in the range between $10^{10} M_\odot$ and $10^{12} M_\odot$, with $X=-2.4$ for $M_h=10^{11} M_\odot$, the latter coinciding with the mean $X$ value in this work.

In the case of Einasto profiles, the comparison with the literature is more difficult since the effects of baryonic feedback are not explicitly reported through its free parameters. For example in \cite{2019MNRAS.483.4086B} they obtain, within (zoom-in) hydro-dynamical simulations and for $\sim \SI{E10}{\Msun}$ halos, a reduction on inner-halo densities (i.e. at the convergence radius $r=\SI{0.26}{\kilo\parsec}$) of up to $\SI{45}{\percent}$ in WDM cosmologies with respect to the analogous WDM-only simulations. A reduction which is about $15 \%$ more pronounced than comparing WDM to CDM only simulations \citep{2019MNRAS.483.4086B}. Even if all the resulting DM halos are there fitted with the Einasto profile, these density reductions are only expressed through its concentration parameter (which is a function of the scale radius and the virial radius). They found indeed that Einasto profiles in WDM cosmologies have smaller concentration parameters than the CDM counterparts. This concentration parameter correlates inversely proportional to the Einasto shape parameter (denoted here as $\kappa$) as originally shown through Figs. 13 and 14 in \citet{2014MNRAS.441.3359D}. Thus, considering that $\kappa\approx 0.2$ correspond to CDM-only cosmologies \citep{2014MNRAS.441.3359D,2017MNRAS.471.3547F}, then larger $\kappa$ values are expected for Einasto profiles having flatter inner-halo slopes \citep{2014MNRAS.441.3359D}. Indeed, our results are qualitatively in line with those reported in \cite{2019MNRAS.483.4086B} since we find a mean value of $\kappa \approx 0.4$ (see \cref{fig:parameter-distribution:einasto}), thus implying smaller concentration parameters than for CDM profiles. However more work from hydro-dynamical simulations using Einasto fitting profiles is needed in order to make a proper quantitative comparison with our statistical results.

We would like also to point out that many galaxies in the SPARC data set are missing significant information in the outer halo (e.g. due to faint stars) or show a complex behavior (oscillatory-pattern) in their rotation curves (see e.g. right box in \cref{fig:benchmark:total-rotation-curves}). In any case, it does not allow to univocally determine the cutoff parameter (i.e. $W_p$) for the fermionic DM model since any sufficiently large $W_p$ would not change the $\chi^2$ value (see e.g. bottom left panel of \cref{fig:chi-analysis}). Nevertheless, there are some other individual galaxies where the escaping particles effects are clearly preferred. Interestingly, all of those galaxies are of magellanic type: NGC0055 (Sm), UGC05986 (Sm), UGC05750 (Sdm), UGC05005 (Im), F565-V2, (Im), UGC06399 (Sm), UGC10310 (Sm), UGC07559 (Im), UGC07690 (Im), UGC05918 (Im) and UGC05414 (Im).

Moreover, many galaxies, which are poorly fitted by any of the considered models, show \textit{short range} oscillations in their rotation curves with more than one maximum. None of the models can provide a clear explanation of that phenomena, found usually in non-magellanic galaxy types: e.g. NGC2403 (Scd), UGC02953 (Sab), NGC6015 (Scd), UGC09133 (Sab), UGC06787 (Sab), UGC11914 (Sab), NGC1003 (Scd), NGC0247 (Sd), UGC08699 (Sab) and UGC03205 (Sab).

On phenomenological grounds, in the fermionic DM model it is possible to vary the width of the maximum bump in the RC through the cutoff parameter in the strong or moderate cutoff regime ($W_p \lesssim 10$). Whether with weak ($W_p \gtrsim 10$) or even without cutoff-effects, the RC solutions of the model show long range oscillations, similar to the isothermal model. In any case, these RC oscillations have a too long wavelength and therefore do not offer a convenient explanation. On the other hand, in the case of strong cutoff, we obtain a narrow maximum bump necessary for many RCs, especially for galaxies of magellanic type (see above for examples), which usually do not show those oscillations, but also for some non-magellanic galaxy types, e.g. NGC5585 (Sd), NGC7793 (Sd), UGC06614 (Sa), ESO079-G014 (Sbc), F571-8 (Sc), NGC0891 (Sb), UGC06614 (Sa), UGC09037 (Scd), NGC4217 (Sb), UGC04278 (Sd).

NFW and Burkert models cannot explain variations of the inner and outer rotation curve because the parameters ($\beta$ and $\gamma$) responsible for such a behaviour (see \cref{eqn:hernquist}) are fixed. Additionally a transition from the inner to the outer halo is generally characterized by $\alpha$ (or $\kappa$). In contrast to NFW and Burkert, the DC14 and Einasto models have a free parameter which affects the inner/outer rotation curve steepness and the sharpness of the halo transition, simultaneously. Such a flexibility is reflected in generally better $\chi^2$ values. Nevertheless, the goodness for oscillating RCs remains rather poor.

\section{Fermionic halos}
\label{sec:fermionic-halos}
This section is mainly devoted to fermionic halos where we include the analysis of the DM surface density relation (SDR), originally based on the Burkert model. Further, we highlight the rich morphology of fermionic DM halos and show that particular solutions can be associated with different DM models. As typical examples, we present detailed rotation curve fits and their corresponding analysis for three selected galaxies indicating the limitations of the observational data and/or the case where data supports for one clear maximum in the rotation curve of each DM model.

\subsection{DM surface density relation}
\label{sec:dark-matter-surface-density}

The constant surface density \citep{2009MNRAS.397.1169D} 
\begin{equation}
\label{eqn:Donato}
	\Sigma_{0D} = \rho_{\rm 0D} r_0 \approx 140_{-50}^{+80} \si{\Msun\per\parsec^2}.
\end{equation} 
is valid for about 14 orders of magnitude in absolute magnitude ($M_B$) where $\rho_0$ is the \textit{central} DM halo density and $r_0$ the one-halo-scale-length --- both of the Burkert model. At $r_0$ the density falls to one-forth of the central density, i.e. $\rho(r_0) = \rho_{\rm 0D}/4$.

Note that the \textit{center} in the Burkert model corresponds to the plateau in the fermionic DM model, i.e. $\rho_{\rm 0D} \approx \rho_{\rm p}$ where the plateau density $\rho_{\rm p}$ is defined at the first minimum in the RC. Following the definition of the Burkert radius $r_0$, we identify the one-halo-scale-length $r_B$ of the fermionic model such that $\rho(r_B) = \rho_p/4$. We thus calculate the product $\rho_p r_B$ for each galaxy.

\begin{figure}%
	\centering%
	\includegraphics[width=\hsize]{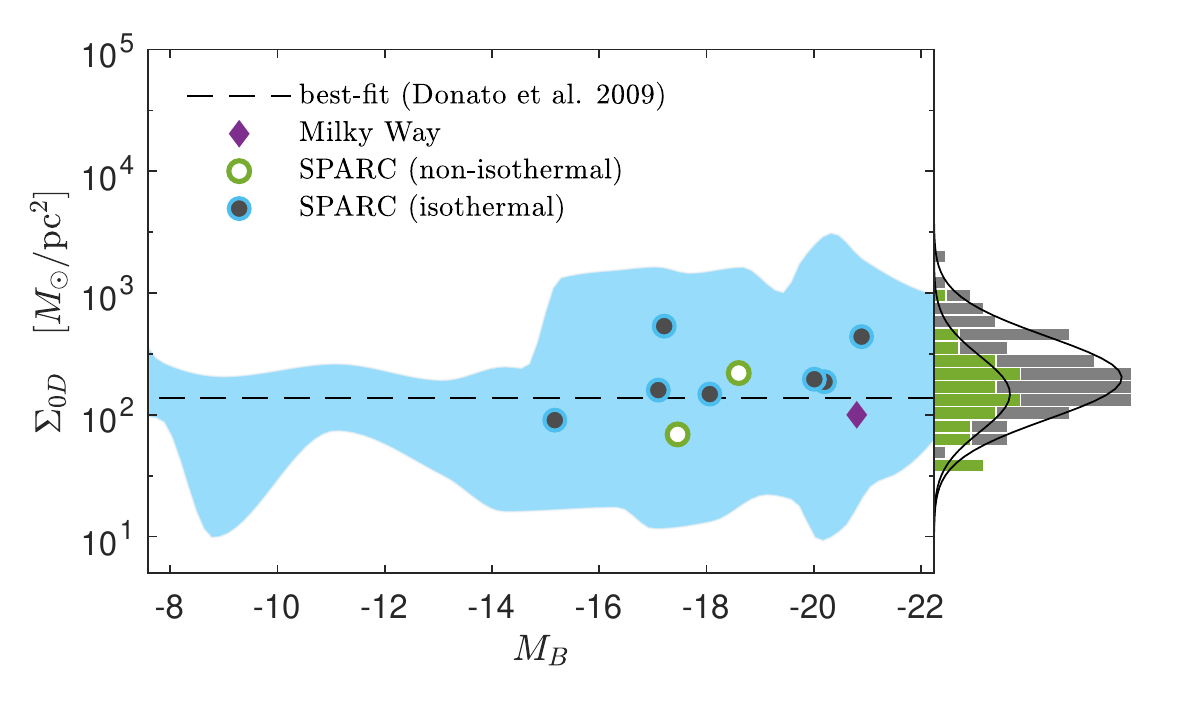}
	\caption{DM surface density ($\Sigma_{0D}$) predictions in the fermionic model using the SPARC data-set. The blue region indicates the delimited area by the $3\sigma$ error bars of all the data points in \citet{2009MNRAS.397.1169D}. The violet diamond represents the MW as analyzed in \cite{2018PDU....21...82A}. The absolute magnitude was taken from the Carnegie-Irvine Galaxy Survey \citep{2011ApJS..197...21H}, providing nine overlapping galaxies (blue and green circles). The full  sample (gray bars) and sub-sample (green bars) including only for non-isothermal solutions ($W_p < 10$), are shown as histograms, both following approximately a Gaussian distribution.}%
	\label{fig:SPARC:Donato}%
\end{figure}

The absolute magnitude $M_B$ was taken from the Carnegie-Irvine Galaxy Survey \citep{2011ApJS..197...21H}, providing nine overlapping galaxies with the SPARC sample. These candidates are well in agreement with the DM surface density observations, see \cref{fig:SPARC:Donato}. The shown candidates include isothermal-like (blue outlined points) and non-isothermal (green circles) solutions. For comparison, the results are amended by the MW solution following the fermionic RAR model \citep{2018PDU....21...82A}.

\def\ROOTPATH{figure/Donato}

Although absolute magnitude information is incomplete, all of the predicted DM surface densities are within the range of the $3\sigma$ area as well. This is visualised by a histogram for the full sample (dark grey bars, 120 galaxies) with comparison to the sub-sample (green bars, 44 galaxies) including non-isothermal solutions (i.e. $W_p < 10$). Considering the sub-sample only, we obtain a mean surface density of about $\SI{148.5}{\Msun/\parsec^2}$, fully inside the 1-$\sigma$ uncertainty in \cref{eqn:Donato}.

The SDR given by \cref{eqn:Donato} is qualitatively consistent with the scaling relation $M_h \sim r_h^2$ as given by \cref{eqn:rel:Mh-rh}. Due to the different halo profiles of the fermionic model, ranging from polytropes of $n=5/2$ to isothermal (see section \ref{sec:morph}), there is a non-linear relation between the halo radius $r_h$ and the one-halo-scale-length $r_B$. Nevertheless, considering that the halo is nearly homogeneous up to approximately the halo radius $r_h$ (i.e. $M_h \sim \rho_p r_h^3$), and that $r_h \sim r_B$ we obtain $\rho_p r_B \approx \const$. (see also \citealp{2019PDU....24..278A}).

\subsection{Morphology of fermionic halos}
\label{sec:morph}

We are interested in providing approximate analytic (and semi-analytic) expressions for the halo morphology associated with the corresponding fermionic solutions.

On halo scales, i.e. for $r \sim r_h$ with $r_h$ being the halo radius, the fermionic DM model resembles the King model \citep{1966AJ.....71...64K}. These DM profiles are characterized by a cored inner halo (i.e. a flat inner slope) followed by a transition towards a finite mass. When applied to galactic halos, such a transition can show different behaviours leading to a rich morphology of density profiles, from polytropic-like to power law-like, mainly depending on the value of $W_p$. Consequently, this cutoff (or particle escape) parameter controls the sharpness of the inner-outer halo transition as well.

Very similar characteristics are also produced by the Einasto model, although with a wider spectrum for the sharpness of the transition described by the parameter $\kappa$.

\begin{figure}%
	\centering%
	\includegraphics[width=\hsize]{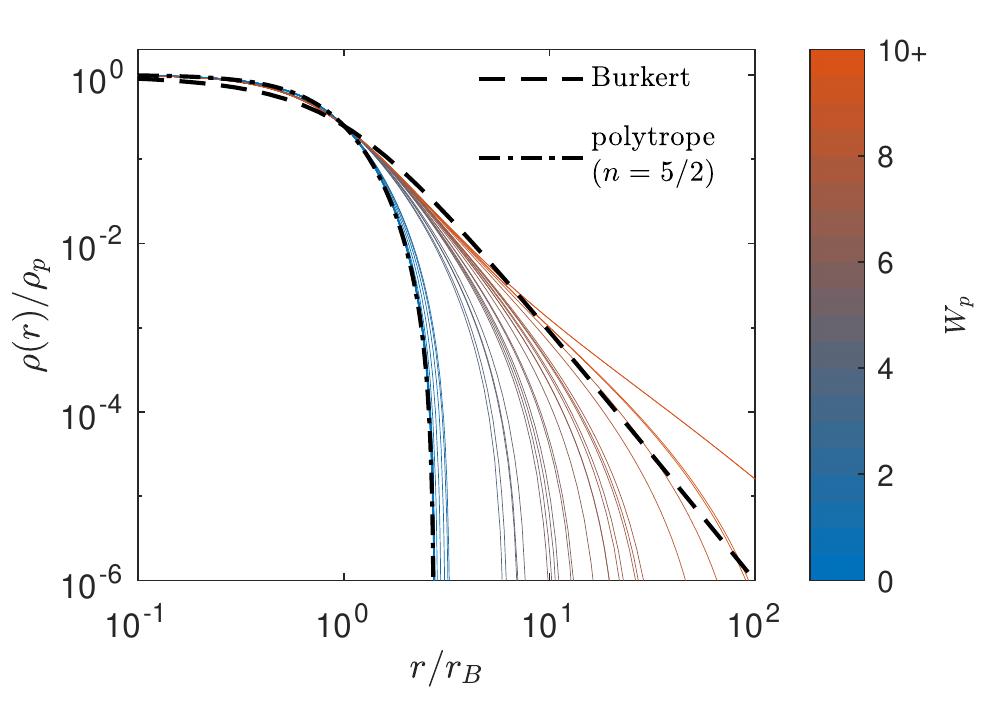}%
	\caption{Morphology of best-fitted fermionic halos (solid). For comparison the Burkert profile (dashed) and a polytrope of $n=5/2$ (dot-dashed) are added. Shown are normalized halo density profiles where the density is normalized by the inner halo density $\rho_p$ and the radius by the Burkert radius $r_B$ fulfilling $\rho(r_B) = \rho_p/4$.}%
\label{fig:halo-profiles}
\end{figure}

In the fermionic model, the sharpness of the transition can be quantified as follows: for $W_p\ll 1$, the fermionic density profiles are polytropes of index $n=5/2$ as clearly shown in \cref{fig:halo-profiles} for bluish solutions (see \citealp{2015PhRvD..92l3527C} for a derivation of the polytropic $n=5/2$ equation and its link with the fermionic solutions, further justifying our results). In contrast, for negligible escape of particles (e.g. $W_p \gtrsim 10$), the fermionic halos become isothermal-like, with $\rho(r) \propto r^{-2}$ down to the virial radius as evidenced through the reddish solutions in the same figure. Finally, for $W_p$ values in between the above limiting cases, the fermionic DM halo profiles start to develop a power-law trend almost matching the Burkert profile for $W_p\approx 7$ (see \cref{fig:halo-profiles}).

These results agree with those obtained in Sec. VII of \citet{2015PhRvD..91f3531C} for the classical King model, which correspond to our fermionic model in the dilute regime (see also paragraph below). There it is shown that the Burkert profile gives a good fit of such King distribution for a value of $W_p$ close to the point of marginal thermodynamical stability, the latter given at $W_p^{(c)} = 7.45$. In \citet{2015PhRvD..91f3531C} this marginal solution is labelled through the equivalent parameter $k$.

Statistically, we identified $W_p \approx 10$ as a discriminator between two groups and further detailed in Appendix \ref{sec:appendix:parameter-distribution}. Even if the precise value might be biased by the data, there is an interesting underlying physical explanation for it: fermionic DM profiles as obtained from a MEP are thermodynamically and dynamically stable for $W_p \leq 7.45$ if they are in the dilute regime ($\theta_0 \ll -1$ where $W_0 \equiv W_p$), while in the core-halo regime ($\theta_0 > 10$) the same stability conditions hold for a bounded $W_p \in [W_p^{\rm min}, W_p^{\rm max}]$. For typical SPARC galaxies with a halo mass of $\sim \SI{5E10}{\Msun}$, we find $W_p^{\rm max}\sim \num{E-2}$.

\def\ROOTPATH{figure/HaloProfilesComparison}

The point of stability-change for the classical King model was first obtained in \cite{2015PhRvD..91f3531C}, and further re-derived here for fermionic DM profiles but this time for realistic average halo-mass galaxies, as obtained from the SPARC data-set, following the thermodynamic analysis of \cite{2021MNRAS.502.4227A}. That is, typical galaxies with halo masses of about $\sim \SI{5E10}{\Msun}$ and with appreciable escape of particles (i.e. $W_p \ll 1$) are thermodynamically and dynamically stable, suggesting a deep link between thermodynamics of self-gravitating systems and galaxy formation. A case-by-case stability analysis in relation to observed galaxies (e.g. SPARC data-set) is out of the scope of the present paper and will be the subject of a future work. 

\looseness=2
\def\ROOTPATH{figure/BenchmarkTotalRotationCurves}\begin{figure*}%
	\centering%
	\includegraphics[width=\hsize]{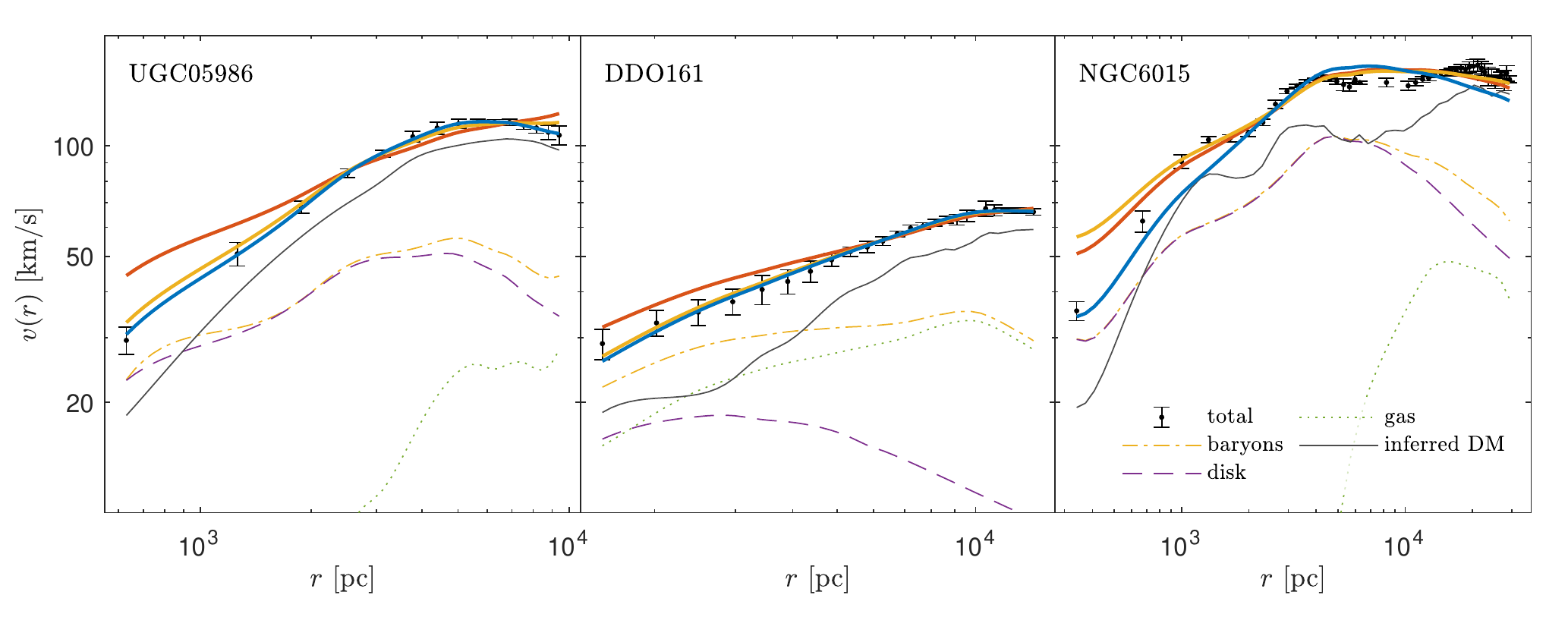}%
	\caption{Total rotation curves and their composition, shown for UGC05986 (left), DDO161 (center) and NGC6015 (right). The thick lines are the best-fits of three competing DM models: Fermionic (blue), DC14 (yellow) and NFW (red). For clarity we have excluded Einasto.}%
	\label{fig:benchmark:total-rotation-curves}%
\end{figure*}
\subsection[Best-fit analysis]{Diversity of SPARC rotation curves}
\label{sec:fermionic-halos:diversity}

We show in this section a detailed $\chi^2$ analysis of the RC fits for three selected galaxies, each representing some characteristics of given observational data. We divide the SPARC galaxies in three groups by the inferred DM component as explained next. This analysis is based on the fermionic model where such a grouping seems to be appropriate to select galaxies with valuable predictions about the inner halo.

\begin{figure}%
	\centering%
	\includegraphics[width=\hsize]{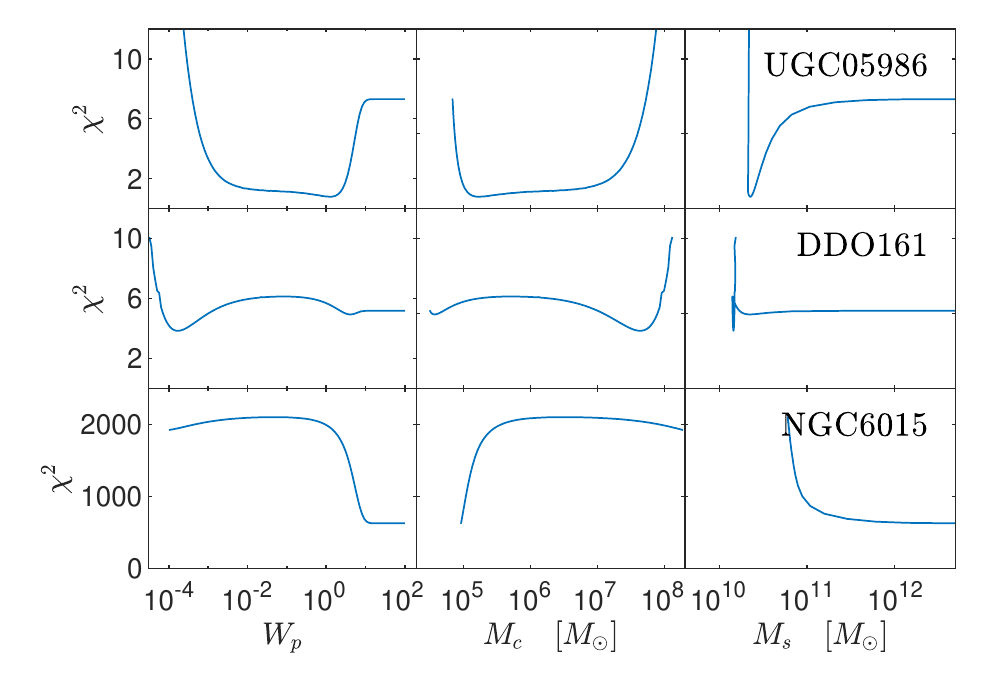}%
	\caption{$\chi^2$ profiles of the fermionic DM model for three benchmark galaxies: UGC05986 (top panels), DDO161 (middle panels) and NGC6015 (bottom panels).}%
	\label{fig:chi-analysis}%
\end{figure}

The first group, represented by UGC05986, shows only a single maximum in its DM RC, i.e. a rising trend in the inner halo followed by a clear turning point, as can be seen by the data points in the left plot of \cref{fig:benchmark:total-rotation-curves}. In the same plot for UGC05986 we show the best-fits of the competing DM models as thick curves, i.e. the fermionic (blue), DC14 (yellow) and NFW (red). Regarding the fermionic model, this kind of RC is better fitted by the solutions with a significant escape of particles ($W_p \lesssim 10$), as can be explicitly seen through the $\chi^2$ valleys in top panels of \cref{fig:chi-analysis}. However, due to the lack of information in the inner halo structures there is some uncertainty in the strength of particle escape. The uncertainty is physically better reflected in the core mass $M_c$ which covers about two orders of magnitude (see middle panel of first row in \cref{fig:chi-analysis}).

\def\ROOTPATH{figure/Chi2Analysis}

This result goes totally in line with an analogous phenomenological analysis \citep{2019PDU....24..278A}, developed for typical dwarf, spiral and elliptical galaxies within the RAR model. According to that analysis (done for $mc^2\approx \SI{50}{\kilo\eV}$), the maximal core mass of larger galaxies is limited by the critical configuration where the quantum core becomes unstable and collapses to a BH of mass $M_c^{cr} \approx \SI{2E8}{\Msun}$.

Among the cases, which are disfavored, are the ones with very large total DM masses $M_s$ corresponding to isothermal-like halos and implying negligible escape of particles ($W_p \gtrsim 10$). These solutions provide a minimal core mass $M_c$ with a huge uncertainty in the total mass.

The second group, represented by DDO161, shows a rising part in the RC towards a maximum without a clear turning point compared to the first group, see central plots in \cref{fig:benchmark:total-rotation-curves}. Fitting those galaxies for different $W_p$ values does not favor solutions with or without escaping particles effects. The variation in the $\chi^2$ value remains rather small, see middle panels of \cref{fig:chi-analysis}.

Finally, the third group, represented by NGC6015, shows some oscillations in the RC, mainly in the outer halo, see right plots in \cref{fig:benchmark:total-rotation-curves}. There are various and speculative reasons for the oscillation, e.g., ongoing merging process, deviation from equilibrium, etc. In any case, those galaxies are clearly better fitted by extended isothermal-like halos ($W_p \gtrsim 10$) --- although being far from good --- see bottom panels of \cref{fig:chi-analysis}. Such solutions provide a wide halo maximum followed by a flat RC. In contrast, the non-isothermal solutions with a cutoff provide only a narrow maximum in the halo, followed by a Keplerian decreasing tail.

It is worth to recall that different DM models such as the fermionic model, NFW, DC14 and others are not appropriate to fit the oscillations, characterized through multiple maxima in the RC. All solutions with a wide halo are suitable to fit the oscillations well on average, although the best-fits remain rather poor, leaving almost no insight into the physical properties of DM on halo scales for those galaxies.

\section{Summary and conclusion}
\label{sec:conclusion}

For the case of disk galaxies, as provided by the SPARC data-set (see section \ref{sec:data}), we have studied the galactic rotation curves and different galaxy scaling relations --- such as the Radial Acceleration Relation, MDAR and DM surface density relation (SDR) --- from an alternative perspective in which the halos are formed through a MEP.

Within this paradigm we considered the DM halo as a self-gravitating system of neutral fermions at finite temperature while the baryonic mass components were provided from the SPARC data-set. 

For comparison, we have taken into account empirical DM fitting models motivated from DM-only simulations like NFW (within CDM) and Burkert (within WDM); the DC14 (or generalized NFW) model which contains different physics such as the influence of baryonic feedback in the morphology of CDM halos; and the Einasto model as recently studied in \citet{2019MNRAS.483.4086B} accounting for baryonic effects (through hydrodynamical zoom-in simulations) in either cosmology. Finally, their best-fits to the acceleration relations and SPARC RCs were compared with the fermionic model (see sections \ref{sec:result:ac} and \ref{sec:result:gof} respectively).

For all competing DM models, we fitted the DM contribution to the RC as inferred from the given total rotation curve and the baryonic component (see section \ref{LM-fitting}).

An alternative fitting approach is minimizing the least square errors of the total rotation curve (e.g. $V_i = V_{i, \rm tot}$) where the predictions are a composition of a theoretical DM halo model and the baryonic component inferred from observation. On theoretical ground, such an approach is not identical to the approach explained in section \ref{sec:fitting} since the propagation of uncertainty produces a somewhat different weighting. However, we have repeated the same analysis on both approaches and obtained consistent results despite few minor numerical variations, and without changing any qualitative conclusion obtained in this work.

The main results of this work can be summarized as follows, according to three different issues.

\subsection{Acceleration relations}
The Radial Acceleration Relation as well as MDAR analyzed here are based on an averaging of many spiral galaxies and hold for different Hubble types. Our analysis shows that all competing DM models are able to reproduce those relations, although without a clear favourite because all are similarly good (see \cref{fig:acceleration:grid}). For instance, for all DM models, we obtain nearly identical values for $\SYMafrak$ as required in \cref{eq:mcgaugh-fit}. This result is in line with a recent analysis done in \citet{2020JCAP...06..027K}. 

\subsection{Individual rotation curve fittings}
A deeper understanding of the Radial Acceleration Relation and MDAR is backed by a goodness of model analysis for $120$ filtered and individual galaxies of the SPARC data set covering different Hubble types. The DM contribution to the RCs reflects some diversity in galaxies which, in general, are better fitted by cored DM halo models instead of cuspy (e.g. NFW, see \cref{fig:goodness:all}). This is not only in agreement with a similar analysis done by \citet{2020ApJS..247...31L}, but totally in line with the results of \citet{2020JCAP...06..027K}. That is, it has been shown here that the DM halo models suitable to explain the acceleration relations do not necessary explain well the SPARC RCs.

Comparing the fitting goodness of the superior DC14 with the inferior (and statistically disfavoured) NFW implies that baryonic feedback mechanism is important in galaxy formation. On the other hand, we found that the fermionic model, compared to DC14 or Einasto, is equally good in fitting galaxies which require a significant escape of particles (i.e. $W_p < 10$, see section \ref{sec:result:gof} and \cref{fig:goodness:with-cutoff}). Those galaxies are characterized by a flat inner halo, justified by very different physical principles in the most favored DM profiles: DC14 and Einasto models rely on complex baryonic feedback processes, while the MEP scenario involves a quasi-thermodynamic equilibrium of the DM particles. This may imply that for those galaxies baryonic feedback is less relevant, thus hinting on the importance of a quasi-thermodynamic equilibrium that may be reached in those DM halos.

\subsection{Fermionic halos}
We found that for SPARC galaxies the \textit{constancy} of SDR, originally based on the Burkert model \citep{2009MNRAS.397.1169D}, is also achievable within the fermionic DM model. From the Carnegie-Irvine Galaxy Survey \citep{2011ApJS..197...21H} we have further extracted the absolute magnitude for nine overlapping SPARC galaxies. The surface density predictions of that sub-sample are well in agreement with observations.

Additionally we demonstrated that particular solutions from the rich morphology of fermionic DM on halo scales can be associated with different empirical DM models, depending on the strength of particle evaporation (described by $W_p$). Fermionic DM halos are polytropic-like (with $n = 5/2$) for strong evaporation ($W_p \ll 1)$,  transition into profiles similar to Burkert for values of $W_p$ close to the point of marginal thermodynamical stability (given at $W_p^{(c)} = 7.45$) and finally develop isothermal tails for negligible evaporation ($W_p \gg 10$).

Interestingly, the mean $\kappa$ value obtained here for the Einasto model is roughly $0.4$, implying a pronounced inner-halo density drop (i.e leading to an almost flat inner-slope, see \cref{fig:profile-illustration-mep}), such that the halos look more like the fermionic profiles.

Indeed, following the work done in \cite{2021MNRAS.502.4227A} for such fermionic profiles, it can be found that typical galaxies belonging to this sub-group (with $W_p \ll 1$) are thermodynamically and dynamically stable, with an outer-halo morphology of polytropic nature (see \cref{fig:halo-profiles}). This may be evidencing a fundamental and deep link between thermodynamics of self-gravitating fermions and galaxy formation and morphology.

\begin{acknowledgments}
    This work was founded by the Consejo Nacional de Investigaciones Científicas y Técnicas (CONICET), grant number 11220200102876CO. We thank B. Famey for useful discussions. We thank the anonymous referee who helped to improve the presentation of this work.
\end{acknowledgments}

\appendix
\section{Parameter correlations of fermionic halos}
\label{sec:parameter-correlations}

We analyzed different pairs of structural galaxy parameters, obtained from core-halo best-fit solutions of the fermionic model for a particle mass of $\SI{50}{\kilo\eV}$.

Of interest here are values at the halo such as the halo radius $r_h$ and halo mass $M_h = M(r_h)$. The halo radius $r_h$ is defined at the second maximum in the rotation curve $v(r)$.

\begin{figure}%
	\centering%
	\includegraphics[width=\hsize]{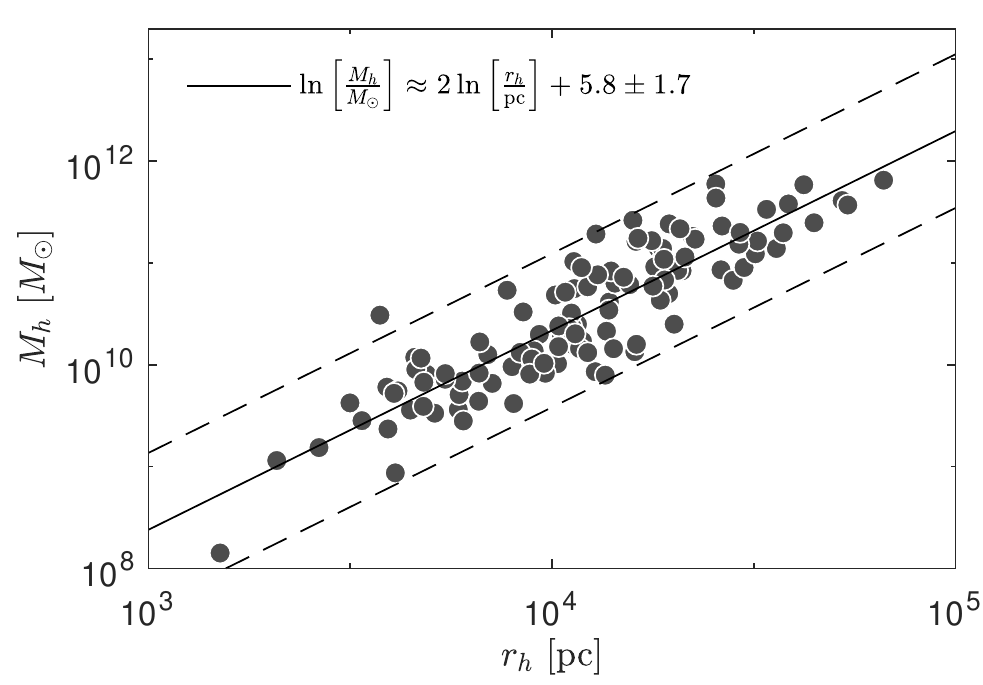}%
	\caption{Halo parameter correlation for the best-fits of the fermionic model. The dashed lines indicate the $\SI{95}{\percent}$ CI.}%
\label{fig:parameter-correlction:core-halo}
\end{figure}

For mass and radius we obtain values located mainly in the intervals $r_h \in [10^3,10^5]\si{\parsec}$ and $M_h \in [10^8,10^{12}] \si{\Msun}$. As shown in \cref{fig:parameter-correlction:core-halo} the halo radius and mass follow a clear relation described by \begin{equation}
	\label{eqn:rel:Mh-rh}
	\ln \qbracket{\frac{M_h}{M_\odot}} \approx 2\ln \qbracket{\frac{r_h}{\mathrm{pc}}} + 5.8 \pm 1.7
\end{equation} Interestingly, this relation is independent of particle escape, i.e it holds for solutions with an isothermal-like halo developing a flat tail as well as for polytropic non-isothermal halos implying a large escape of particles.

\def\ROOTPATH{figure/ParameterCorrelations}

\section{Parameter distributions}
\label{sec:appendix:parameter-distribution}

\def\ROOTPATH{figure/ParameterDistribution_MEPP}\begin{figure}
	\centering%
	\includegraphics[width=\hsize]{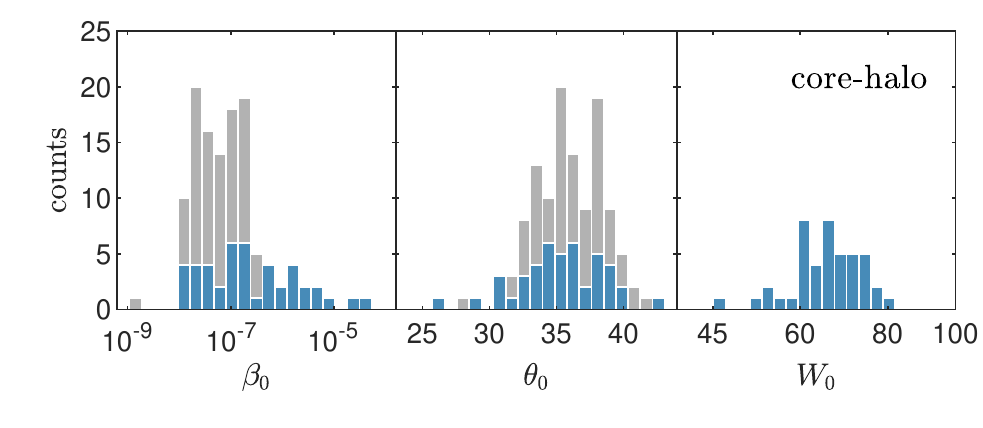}
	\includegraphics[width=\hsize]{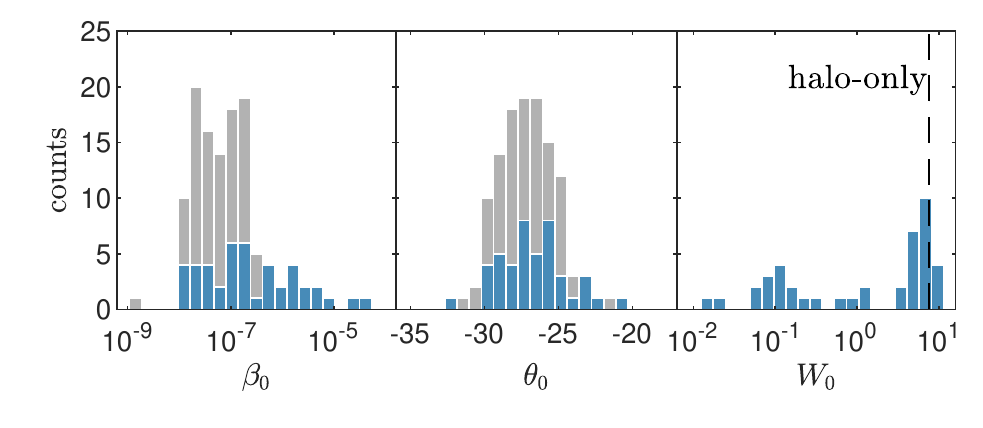}
	\caption{Distribution of the best-fit parameters for the fermionic model with a \textit{core-halo} (top) and fully diluted (bottom). The blue bars represent the sub-sample of non-isothermal solutions (44 galaxies) with $W_p \lesssim 10$ (top) and $W_0 \lesssim 10$ (bottom), respectively, while the gray bars include the full sample (120 galaxies). Note that the values at plateau of the core-halo solutions (i.e. $\beta_p$, $\theta_p$, $W_p$) can be identified with the values at the center of the halo-only solutions (i.e. $\beta_0$, $\theta_0$, $W_0$). In the case of halo-only solutions (bottom) the dashed line represents the stability change where fermionic DM halos become unstable for $W_p\equiv W_0 \gtrsim 7.45$. \citep{2015PhRvD..91f3531C}}%
	\label{fig:parameter-distribution:mepp}%
\end{figure}
\def\ROOTPATH{figure/ParameterDistribution_DC14}\begin{figure}
	\centering%
	\includegraphics[width=\hsize]{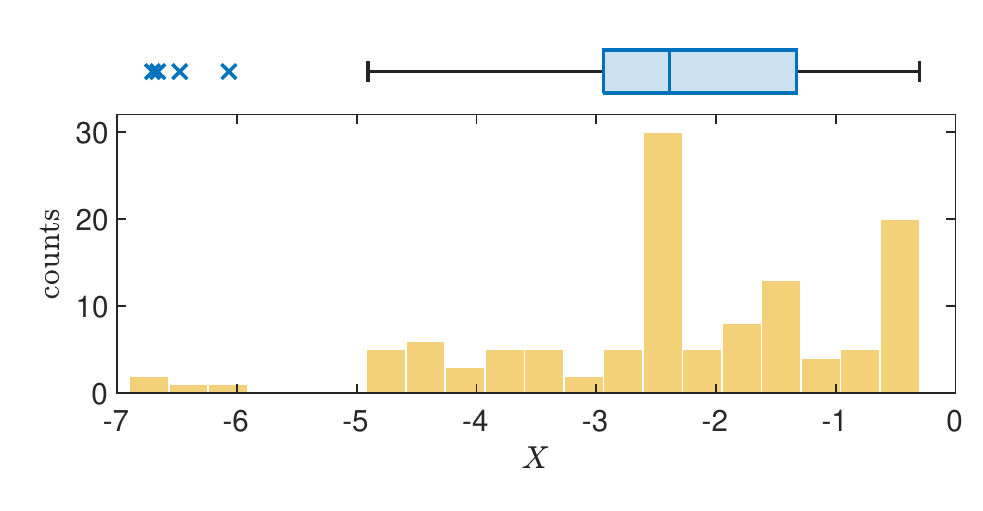}
	\includegraphics[width=\hsize]{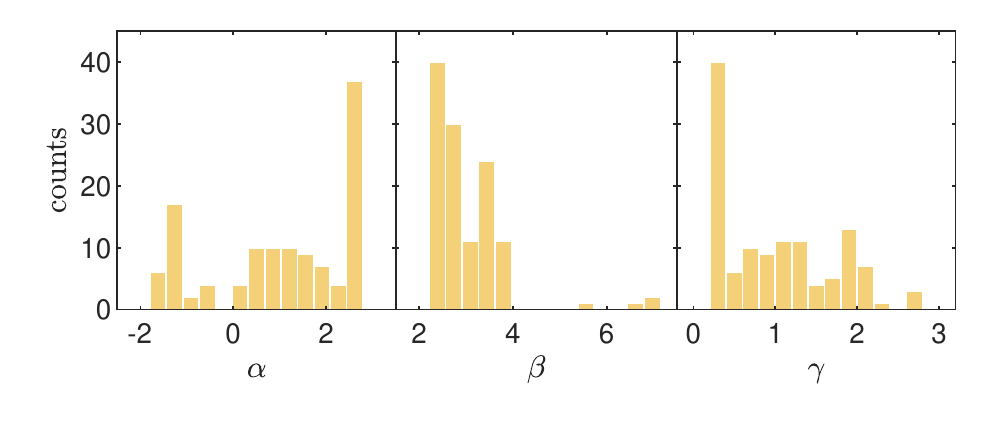}
	\caption{Distribution of the best-fit parameters for the DC14 model. Above the histogram a boxplot is shown with a median value at $X \approx -2.4$. The mean value is very close to the median value. The corresponding values for $\alpha$, $\beta$ and $\gamma$ are calculated from the best-fitted $X$ value following \cref{eqn:dc14:alpha,eqn:dc14:beta,eqn:dc14:gamma}.}%
	\label{fig:parameter-distribution:dc14}%
\end{figure}
\def\ROOTPATH{figure/ParameterDistribution_Einasto}\begin{figure}
	\centering%
	\includegraphics[width=\hsize]{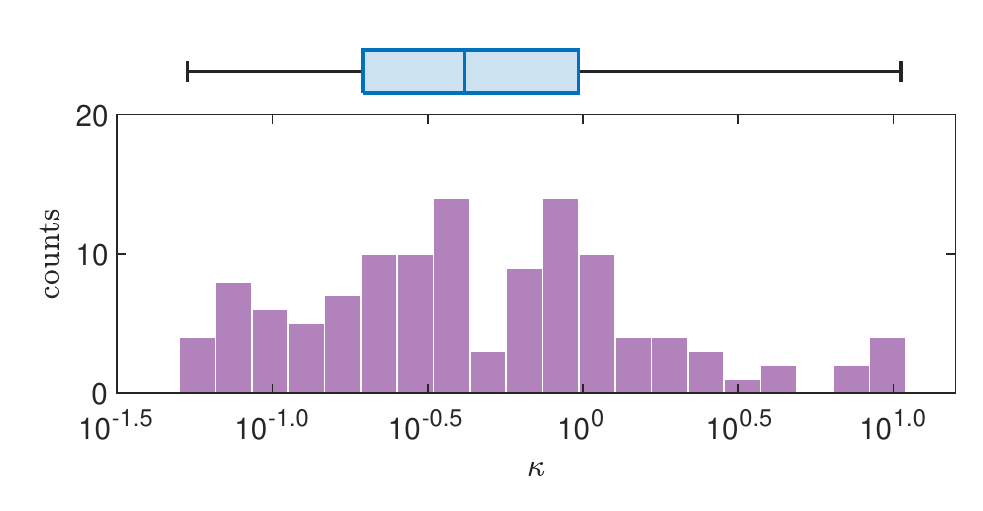}
	\caption{Distribution of the best-fit parameter for the Einasto model. A boxplot (above the histogram) is shown with a median value of $\kappa \approx 0.42$ and mean value $\kappa \approx 0.46$.}%
	\label{fig:parameter-distribution:einasto}%
\end{figure}

We analysed the distribution of configuration parameters which affect the shape of the DM RC for each DM model. On this basis, we are interested only in fermionic, DC14 and Einasto models, which are the only ones where the halo slopes (usually characterized by $\gamma$) depend on one configuration parameter.

Starting with the core-halo solutions, see upper plots in \cref{fig:parameter-distribution:mepp}, we find that the majority of central temperature values falls in the range $\beta_0 \in [\num{E-8}, \num{E-6}]$ corresponding to solutions in which the central DM cores are far from reaching the core-collapse towards a SMBH \citep{2019PDU....24..278A,2021MNRAS.502.4227A}. The distribution of the central degeneracy values looks like a Gaussian with the mean value at about 36, far in the degenerate regime $\theta_0 \gtrsim 10$. The majority is in the range $\theta_0 \in [30, 40]$. A similar distribution pattern is given for $W_0$. However, for core-halo solutions (i.e. $\theta_0 \gtrsim 10$) it is better to look at the plateau cutoff $W_p$ which can be identified with $W_0$ of a corresponding halo-only solution.

The plateau of a core-halo solution acts as a connection between the halo and the embedded core. Therefore, every core-halo solution of the fermionic model has a corresponding halo-only solution describing the diluted halo without the embedded, degenerate core. The corresponding values for such halo-only solutions are given at the plateau which resemble the inner halo, see lower plots in \cref{fig:parameter-distribution:mepp}. The plateau temperature shows a very similar distribution due to the tiny temperature changes when in the low temperature regime ($\beta_0 \ll \num{E-4}$) where pressure effects are negligible leading to $\beta_p \approx \beta_0$. The plateau degeneracy distribution looks also similar to the central degeneracy but being mirrored and shifted to the negative (diluted) regime. We find the relation $\theta_p \approx -0.7 \theta_0 - 1.2$.

Of great interest is the plateau cutoff $W_p$ which is a proxy for the central cutoff $W_0$ parameter and provides better insights about the halo shape. The plateau cutoff describes the particle escape intensity on halo scales and is defined as $W_p = W(r_p)$ with the plateau radius $r_p$ located at the first minimum in the DM RC. The lower $W_p$ the more truncated is the halo due to evaporation. See also section \ref{sec:morph} for a discussion about the halo morphology and comparison with other DM models.

Within the fermionic model we identify two groups in the $W_p$ distribution of core-halo solutions and $W_0$ distribution of halo-only solution, respectively: (1) an isothermal (non-truncated) group with 76 galaxies and (2) a non-isothermal (truncated) group with 44 galaxies. Both groups are divided at about $W_p \approx 10$ (core-halo) and $W_0 \approx 10$ (halo-only), respectively.

For the first group the outer halo seems to be isothermal (i.e. characterized by a flat tail and $\rho(r) \propto r^{-2}$ for large enough $r$). However, for those galaxies the exact value of $W_0$ cannot be determined due to insufficient and/or too limited information in the outer halo data, and thus they are not shown in \cref{fig:parameter-distribution:mepp}. In contrast, the second group seems to have a broader data coverage and/or admitting for a cleaner description of the outer RC, allowing for a better constrain of the outer halo of the fermionic DM profiles (i.e. with finite $W_p \lesssim 10$ and consequent non-isothermal halo tails). See \cref{fig:halo-profiles} for a comparison between typical solution of both groups.

For halo-only solutions (i.e. $\theta_0 \lesssim -5)$ there is an interesting accumulation of galaxies around the particular value of $W_0 \approx 7.45$, see dashed line in \cref{fig:parameter-distribution:mepp}. This particular value reflects the point of stability change where the diluted solutions become unstable, i.e. for $W_0 \gtrsim 7.45$ \citep{2015PhRvD..91f3531C}, and may indicate physical insights into galaxy formation. Interestingly, for this value of $W_0$ the density profile of the classical King model resembles the Burkert profile (see section \ref{sec:morph}). 

In the case of the DC14 model the majority of best-fitted $X$ values is in the range $[-5, 0]$ with a peak at the median of $X \approx -2.4$ as indicated by the boxplot in \cref{fig:parameter-distribution:dc14}. The crosses may be outliers. When we calculate the parameters $\alpha$, $\beta$ and $\gamma$ with \cref{eqn:dc14:alpha,eqn:dc14:beta,eqn:dc14:gamma}, there seems to be a grouping in the distribution of $\alpha$ with a separation at $\alpha = 0$, see \cref{fig:parameter-distribution:dc14}. However, \cref{eqn:hernquist} is not defined for $\alpha = 0$. Phenomenologically, $\alpha$ describes the transition from the inner to the outer halo. The larger $\abs{\alpha}$ the more extended is the transition, characterized by a long wide maximum. $\beta$ describes the slope in the outer halo while $\gamma$ describes the slope in the inner halo. For $\gamma = 0$ the DM profiles become cored.

The Einasto model has only a single configuration parameter $\kappa$ describing the transition from the cored inner halo to the outer halo. The larger $\kappa$ the less extended is the RC maximum. As shown in \cref{fig:parameter-distribution:einasto} the majority has $\kappa < 1$ that corresponds to a rather extended RC maximum. This distribution represents well the majority of SPARC galaxies showing an extended outer halo trend without a clear maximum in the outer RC.

\bibliographystyle{aasjournal}
\bibliography{bib-articles,bib-books}
\end{document}